\documentclass[aps,pra,
superscriptaddress,
reprint,twocolumn,preprintnumbers,
amsmath,amssymb,
nofootinbib]{revtex4-1}

\usepackage[titletoc,toc,title]{appendix}

\usepackage{braket}
\usepackage[final]{feynmp}
\usepackage{ifpdf}
\usepackage{comment}
\usepackage{graphicx}
\usepackage{dcolumn}
\usepackage{bm}
\usepackage{hyperref}
\usepackage[titletoc,toc,title]{appendix}
\usepackage{amsmath}
\usepackage{mathrsfs,mathtools}
\usepackage{color}
\usepackage{enumerate}

\makeatletter
\def\endfmffile{%
	\fmfcmd{\p@rcent\space the end.^^J%
			end.^^J%
			endinput;}%
	\if@fmfio
		\immediate\closeout\@outfmf
	\fi
	\ifnum\pdfshellescape=\@ne
		\immediate\write18{mpost \thefmffile}%
	\fi}
	
\def\rf#1{(\ref{#1})}

\newcommand{\beq}{\begin{equation}}
\newcommand{\eeq}{\end{equation}}
\newcommand{\p}{\partial}
	
\makeatother

\begin{document}

\graphicspath{{./Figures/}}


\title{One-dimensional model of chiral fermions with Feshbach resonant interactions}

\author{Abhinav Prem}
\email{abhinav.prem@colorado.edu}
\affiliation{
Department of Physics and Center for Theory of Quantum Matter,
University of Colorado, Boulder, Colorado 80309, USA
}
\author{Victor Gurarie}
\affiliation{
Department of Physics and Center for Theory of Quantum Matter,
University of Colorado, Boulder, Colorado 80309, USA
}

\begin{abstract}
We study a model of two species of one-dimensional linearly dispersing fermions interacting via an s-wave Feshbach resonance at zero temperature. While this model is known to be integrable, it possesses novel features that have not previously been investigated. Here, we present an exact solution based on the coordinate Bethe Ansatz.
In the limit of infinite resonance strength, which we term the strongly interacting limit, the two species of fermions behave as free Fermi gases. In the limit of infinitely weak resonance, or the weakly interacting limit, the gases can be in different phases depending on the detuning, the relative velocities of the particles, and the particle densities. When the molecule moves faster or slower than both species of atoms, the atomic velocities get renormalized and the atoms may even become non-chiral. On the other hand, when the molecular velocity is between that of the atoms, the system may behave like 
a weakly interacting Lieb-Liniger gas. 

\end{abstract}

\maketitle


	\section{Introduction}
	\label{intro}
	
Recent advances in the cooling and trapping of ultra-cold atomic gases, coupled with the ability to tune interactions via Feshbach resonances~\cite{Regal2003,Inouye2004,Stan2004,Gunter2006,Best2009}, have made accessible the regime of strong correlations in many body physics. In particular, it has become possible to confine quantum gases to effectively one-dimensional (1D) geometries, where the effects of strong correlations are enhanced. The experimental realization of strongly interacting 1D systems, such as the Tonks-Girardeau gas~\cite{Paredes2004,Kinoshita2004}, the super-Tonks-Girardeau gas~\cite{Haller2009}, and Fermi gases close to a Feshbach resonance~\cite{Gunter2005,Esslinger2005,Liao2010}, thus provides exciting new opportunities for studying the many-body physics of 1D quantum gases. Recently, considerable theoretical attention has also been paid to 1D Bose-Fermi mixtures, since these are realizable through current cold atom experiments~\cite{Das2003,Yin2012,Batchelor2005,Guan2008,Demler2006a,*Demler2006b,Mathey2004,Cazalilla2003,Giorgini2002,Gritsev2010}.

One such experimentally relevant system is that of a 1D spin-$1/2$ gas of fermions, where it is now routine to tune the atoms forming the gas close to a Feshbach resonance, which induces controllable interactions amongst the fermionic atoms. This system of fermionic atoms, confined to a 1D geometry and interacting via an $s$-wave Feshbach resonance, can be described by the following Hamiltonian,
	\begin{align}
	\label{eqn:01}
	H_F = & \int \, dx \, \Bigg[ \sum\limits_{\sigma=\uparrow,\downarrow} \hat{a}_{\sigma}^{\dagger} \left(-\frac{\partial^2_x}{2 m_a}\right) \hat{a}_{\sigma} + \hat{b}^{\dagger}\left( -\frac{\partial^2_x}{2 m_b}  \right)  \hat{b} \nonumber \\ & + \epsilon_0 \hat{b}^{\dagger} \hat{b} +  g \big(\hat{b}^\dagger \hat{a}_\uparrow \hat{a}_{\downarrow} + h.c. \big) \Bigg],
	\end{align}
where $\hat{a}_{\sigma}^\dagger$, $\hat{a}_{\sigma}$ are fermionic creation and annihilation operators for the open-channel atoms (here $\sigma$ refers to
two hyperfine states of these atoms, labelled by $\uparrow$ and $\downarrow$), $\hat{b}^\dagger$, $\hat{b}$ are those for the closed-channel bosonic molecules
formed by the atoms, $m_a$ and $m_b$ are their respective masses, $\epsilon_0$ is the bare detuning---which can be changed in an experiment by varying the
magnetic field applied to the atoms---and $g$ is the atom-molecule interconversion strength. 
This model, which first appeared in the context of high temperature superconductivity (in three spatial dimensions)~\cite{Lee1989a, *Lee1989b}, is usually referred to as ``the boson-fermion resonance" or ``two-channel" model, and describes the BCS-BEC crossover in fermionic paired superfluids~\cite{Holland2001,Timmermans2001,Gurarie2007,Recati2005}. 

In actual experiments, the parameter $g$ is usually very large, a situation which is typically referred to as the broad resonance regime. Under the assumptions of a large interconversion strength $g$, it is possible to integrate out the molecules to arrive at a purely fermionic Hamiltonian,
\begin{equation} \label{eq:ocm}  \hat H = \int dx \left[ \sum_{\sigma=\uparrow, \downarrow} \hat{a}_\sigma^\dagger \left(- \frac{\partial^2_x }{2m_a} \right) \hat{a}_\sigma - \frac{g^2}{\epsilon_0} \hat{a}^\dagger_{\downarrow}
\hat{a}^\dagger_{\uparrow} \hat{a}_\uparrow \hat{a}_\downarrow
\right].
\end{equation}
This is the familiar and well studied model of two species of fermions interacting via a short range interaction, where the strength of this interaction can be controlled by tuning the parameter $\epsilon_0$. 

An attractive feature of the resulting one-channel model Eq.~\eqref{eq:ocm} is that it is exactly solvable by the Bethe Ansatz technique~\cite{Bethe1931}. First found independently by M. Gaudin~\cite{Gaudin1967a,Gaudin1967b} and by C.N. Yang~\cite{Yang1967} in 1967 as an abstract exercise, the exact solution for this model is now widely used to describe the experiments on 1D fermionic gases interacting via a Feshbach resonance~\cite{Fuchs2004,Tokatly2004}.

However, it is in principle possible to conduct experiments in the case where $g$ is not large, usually termed the narrow resonance regime. In this regime, the physics of these 1D Bose-Fermi mixtures is described by 
the full model Eq.~\eqref{eqn:01} and can well be richer than that of the short ranged interacting fermions described by Eq.~\eqref{eq:ocm}. 

Unfortunately, the model given by Eq.~\eqref{eqn:01} is not integrable and cannot be solved exactly~\cite{Gurarie2006}. This model is however amenable to analysis via the bosonization technique, a program which was carried out in Refs.~\cite{Sheehy2005, Citro2006}. This analysis indeed uncovers additional phases which fermions confined to a 1D geometry and interacting via a narrow Feshbach resonance can form, as compared to the simpler case of a broad resonance. 

However, bosonization is an approximate technique and its application to any particular problem requires making certain assumptions about the possible phases one expects the system to form. It is therefore of considerable interest to have a model of Feshbach resonance which would be exactly integrable in 1D space and thus be exactly solvable, without having to rely on any assumptions. 

Such an integrable, and thus exactly solvable, model of narrow Feshbach resonance in 1D space indeed exists in the literature. Called the 
Quantum Three Wave Interaction (Q3WI) model, it differs from Eq.~\eqref{eqn:01} in that the particles it describes are chiral and move just in one direction. Its Hamiltonian is given by
 \begin{multline}
	\label{q3wi}	
	\hat{H} = \int\,dx \, \Bigg[\sum\limits_{\sigma=1,2} \hat{a}^{\dagger}_{\sigma} \left(-i v_{\sigma} \partial_x \right) \hat{a}_{\sigma} + \hat{b}^{\dagger} \left(-i u \, \partial_x \right) \hat{b}\\ + \nu\,  \hat{b}^\dagger(x) \hat{b}(x) + g \left(\hat{b} \, \hat{a}^{\dagger}_{1} \hat{a}^{\dagger}_{2} + h.c.  \right) \Bigg],
	\end{multline}
where the $\hat{a}_{1,2}$ are fermionic annihilation operators for atoms and $\hat{b}$ is that for the bosonic molecules, $v_1$ and $v_2$ are the velocities of the chiral fermions, while $u$ is that of the molecule. $\nu$ is the detuning and $g$ the interconversion strength; both are parameters that control the strength of the resonance. Note that the single particle spectrum of Eq.~\rf{q3wi} is unbounded from below. However, as is usual, in our study we will assume a certain cut-off for the allowed values of particle momenta. In its presence, Eq.~\rf{q3wi} has a well defined ground state.

This model was shown to be integrable in Refs.~\cite{Wadati1984a,Wadati1984b} and while the thermodynamic behavior was studied in Ref.~\cite{Ohkuma1985}, the authors of that work did not uncover the different regimes that appear in this model.

Here, we exploit the exact solubility of the Q3WI model through the Bethe Ansatz and study in detail its behavior in the thermodynamic limit. In particular, we analytically study the strongly ($g \to \infty$) and weakly ($g \to 0$) interacting limits of this model at zero temperature. 
Interestingly, we find that in the latter limit this system can undergo various phase transitions and, depending on the model parameters, can form unconventional phases such as one where the system remains non-interacting deep within the Fermi sea or one where the density of fermions diverges in the limit where the coupling $g$ goes to zero,
similarly to the weakly
interacting Lieb-Liniger gas. 

Before summarizing our main results, we first address the feasibility of realizing such a chiral model. The Q3WI model Eq.~\eqref{q3wi} can be obtained from more realistic non-chiral models similar to Eq.~\eqref{eqn:01} if the fermion momenta are restricted to lie in the vicinity of certain Fermi points. 
One simple way to generate this model is to either consider fermionic atoms in a single hyperfine state or to consider spin-polarized fermions. Suppose that all atomic states with their momenta restricted to lie in the interval $k_{f1} < k < k_{f2}$ are filled. Further, suppose that this gas is close to a $p$-wave Feshbach resonance, which can turn pairs of fermions into bosons. Such a system is described by the atom-molecule interconversion term
\begin{align}
H_p = \int dx \left[ \hat a^\dagger \left(- \frac{\partial^2_x}{2m_a}\right) \hat a + \hat b^\dagger \left(- \frac{\partial^2_x}{2m_b} \right) \hat b  \right. \nonumber\\
 \left. + g_p \int dx \, \left( \hat b \, \left( \hat a^\dagger \partial_x \hat a^\dagger  -  \partial_x \hat a^\dagger  \hat a^\dagger \right) + h. c. \right) \right].
\end{align}
Expanding the operator $\hat a^\dagger$ in Fourier series and concentrating on the vicinity of momenta $k_{f1}$ and $k_{f2}$, we reproduce the 
interaction term in Eq.~\eqref{q3wi}, with indices $1$ and $2$ referring to the modes of $\hat a^\dagger$ in the vicinity of the momenta $k_{f1}$ and $k_{f2}$ respectively. Note that `forward scattering,' the terms where all momenta are restricted to the vicinity of one of the two points $k_{f1}$ or $k_{f2}$, will be suppressed when the momentum  transferred to $\hat b$ is small (that is, they will be proportional to the momentum transfer). 

The kinetic energy in Eq.~\eqref{q3wi} is then obtained by linearising the momenta in the vicinity of the Fermi points $k_{f1}$ and $k_{f2}$, as well
as linearising the molecular momenta in the vicinity of $k_{f1}+k_{f2}$. In the continuum, we have that $m_b = 2 m_a$, in which case the molecular velocity is fixed by that of the atoms, $u = (v_1 + v_2)/2$. While our solution for the Q3WI captures this regime (see Sec.~\ref{condensate}), this setting is too restrictive and does not allow us to explore the entire parameter space.

More generally, we can consider carrying out a similar procedure for the full two-channel model Eq.~\eqref{eqn:01}, both in the continuum (see App.~\ref{app:B}) and on a lattice (see App.~\ref{app:C}). Since on a lattice the effective masses of the atom and molecules need no longer be coupled, we find that linearising the lattice two-channel model leads to the Q3WI model with velocities $u,v_1,v_2$ and detuning $\nu$ that can be tuned independently. Hence, this procedure allows us to access all of the regimes of the Q3WI model that we find. 

\paragraph*{\bf{Summary of main results:}}
In this paper, we use the coordinate Bethe Ansatz to find the exact ground state for the Q3WI Hamiltonian, given by Eq.~\eqref{q3wi}. We find that, as is usual in the Bethe Ansatz, the ground state is parametrized by the momenta of the atoms; the molecules, however, only enter the problem through their velocity $u$ and the detuning $\nu$. In the thermodynamic limit, we derive a set of coupled integral equations for the excitation spectrum of the system. In general, these can only be solved numerically, except in the strongly and weakly interacting limits, which we proceed to study. For strong interactions, the system reduces to that of non-interacting Fermi gases, in a way similarly to strongly interacting Fermi or Bose gases in 1D~\cite{LiebLiniger,Lieb1963}.

In the more interesting limit of weak interactions, we find that the nature of the solution is primarily determined by the speed of the molecules relative to that of the atoms i.e., the case where $u$ is greater or less than both $v_1$ and $v_2$ differs qualitatively from the case where $v_1>u>v_2$. In the former case, we find that while the excitation spectra of the atoms remain gapless piece-wise linear functions of momenta, there are certain parameter regimes where the fermions display unconventional behavior. For instance, they can remain interacting deep within the Fermi sea---all the way down to a large momentum cut-off (imposed since the spectrum is unbounded)---or the chirality of the slower moving atoms can change due to the weak interactions.

In the latter case (where $v_1>u>v_2$), we find that the system, in the limit of weak coupling, behaves similarly to the weakly interacting Bose gas. The fact that fermions can behave in this manner is not by itself surprising---it has been known for quite some time that 1D spinless
fermions interacting via a specially chosen two-body potential can be mapped exactly onto an interacting Bose gas, and in particular, can be in its weakly interacting regime~\cite{Shigehara1999, Girardeau2005, Glazman2007}. This is similar to the regime our model displays in the case when $v_1>u>v_2$. We term this the ``quasi-condensate" regime of the Q3WI model, in which the spectrum is no longer
linear but is still gapless. 
The wide variety of behaviors exhibited by the Q3WI model reflects the fact that even weak interactions in 1D may significantly alter the nature of the ground state and of low lying excitations.

In all cases but one (where $v_1 > 0 > v_2$ and $u$ lies between the two) we are able to construct the leading order solution for the ground state densities and excitation spectra, from which we calculate the compressibilities. Being discontinuous as functions of the densities, they indicate that the system undergoes quantum phase transitions. In all cases, the transition is from a phase where, near the Fermi momentum, the system goes from being effectively interacting to non-interacting (or vice-versa).

The rest of the paper is organized as follows: In Sec.~\ref{bethe}, we construct the exact many-body wave-function for the Q3WI Hamiltonian using the coordinate Bethe Ansatz and argue that there are no bound states in this problem. We then derive the Bethe Ansatz equations by imposing periodic boundary conditions. In Sec.~\ref{TBA}, we take the thermodynamic limit of these equations and derive a set of of coupled integral (known as the Thermodynamic Bethe Ansatz (TBA)) equations that describe the ground state densities and excitation spectra of the system at zero temperature. In Sec.~\ref{strongint}, we consider the limit of infinite interaction strength.

In Sec.~\ref{weakint}, we discuss the case of weak interactions qualitatively. In Sec.~\ref{linear} we describe the regime where the molecular velocity is greater than or less than that of the atoms ($u <$ or $> v_1, v_2$). Here, we keep only the leading order, $\delta$-function term in the integral kernel for the TBA equations in order to derive the leading order solution for the ground state densities and excitation spectra in the $T\to0$ limit. Moreover, we show that in all possible regimes, the system undergoes a quantum phase transition.

In Sec.~\ref{condensate}, we describe the case where $v_1>u>v_2$. In this case, as the densities diverge in the limit where $g$ goes to zero, we have to consider the next order term in the interaction strength $g$. For the case where atomic velocities have the same sign, we find the leading order solution for the ground state densities at T = 0 and derive the excitation spectra. As in the previous case, we find that the system exhibits a quantum phase transition. 

	
	\section{Bethe Ansatz for the Q3WI}
	\label{bethe}
	
The Hamiltonian for our model is given by Eq.~\eqref{q3wi}. 
In this model, as is always the case in the presence of a Feshbach resonance, the total number of fermions, $N_f$, is not conserved but the total atomic numbers
\begin{equation}
\label{eq:n1n2}
N_1 = N_{f1} + N_b, \ N_2 = N_{f2}+N_b,
\end{equation} where
		\begin{equation}
		N_{f\sigma}= \int dx \, \hat a_{\sigma}^\dagger(x) \hat a_{\sigma}(x), \,  N_b=   \int dx \, \hat b^\dagger(x) \hat b(x),
		\end{equation}
are conserved. We note that in what follows, we will always work in the regime where $v_1 \neq v_2$ i.e., we will never allow the particles to have equal velocities (the case of equal velocities $v_1=v_2$ is highly degenerate and in principle could be considered separately). 

Consider the following wave function in the sector where $N_1=1$, $N_2=1$,
	\begin{equation} \label{eq:ne2}
	\ket{\Psi}_{1,1} = \int dx\,dy\, \psi(x,y)  \hat{a}^{\dagger}_{1}(x) \hat{a}^{\dagger}_{2}(y)\ket{0} + \int dR\, \phi(R) \hat{b}^{\dagger}(R) \ket{0},
	\end{equation} where $\left| 0 \right>$ denotes the Fock vacuum.
This leads to the following Schr\"odinger equation

	\begin{equation}
	- i \left( v_1 \frac{\partial}{\partial x} + v_2 \frac{\partial}{\partial y} \right) \psi(x,y) + g \phi(x) \delta{(x-y)} = E \psi (x,y),
	\end{equation}
	
	\begin{equation}
	\left(\nu - i u \frac{\partial}{\partial R}\right) \phi(R) + g \psi(R,R) = E \phi(R).
	\end{equation}
	The linear spectrum only allows for a solution in the form
	\begin{equation}
	\psi(x,y) = (\theta{(y-x)} + S \theta{(x-y)}) e^{i k_1 x + i k_2 y} ,
	\end{equation}
where $\theta(x)$ is the usual Heaviside function with $\theta(0) = \frac{1}{2}$. It is worth noting that due to the linear free particle spectrum, the wave function is a discontinuous
function of the position variables at $g\neq 0$. This is in contrast with the usual case of a particle with a conventional kinetic energy---quadratic in momentum---moving in a delta-function potential, where the wave function is continuous. 

Solving these equations, with $E = k_1 v_1 + k_2 v_2$, we find
	\begin{equation}
	\label{eqn:10}
	\phi(x) = \frac{g (1 + S)}{2 (E + \nu - u(k_1 + k_2))}  e^{i (k_1 + k_2) x} ,
	\end{equation}
from which we find the $S$-matrix
	\begin{equation}
	\label{eqn:11}
	S(k_1,k_2) =  \frac{\frac{-i g^2}{2(v_1 - v_2)} + v_1 k_1 + v_2 k_2 - u (k_1 + k_2) - \nu}{\frac{i g^2}{2(v_1 - v_2)} + v_1 k_1 + v_2 k_2 - u (k_1 + k_2)-\nu} .
	\end{equation}
As required by unitarity, $S S^{*} = 1$. We also see that $\lim_{k_1\to k_2} S = -1$, i.e., at low energies (relative momentum, $q = (k_1 - k_2)/2 \to 0$), the fermions undergo total reflection. This is in accordance with a general result stated in~\cite{Gurarie2006}. Note that we can alternatively derive the same $S$-matrix by using the $T$-matrix formalism (see App.~\ref{app:A}). The $T$-matrix formalism has the advantage of automatically resolving the ambiguity built into the choice of the value for $\theta(0)$. 

If we could find $k_1 = q_1 + i\epsilon_1$ and $k_2 = q_2 - i\epsilon_2$, with $\epsilon_{1,2}>0$, such that $S = \infty$, then there would exist bound states in this model. However, we need $v_1 \epsilon_1 = v_2 \epsilon_2$ in order to keep the energy of the state real, and demanding that $\phi(x)$, given by Eq.~\eqref{eqn:10}, be normalizable imposes the condition $\epsilon_1 = \epsilon_2$. Thus, since we are considering the case where $v_1 \neq v_2$, 
there exist no bound states in our problem.

Until the end of this section, let us label the momenta of the $v_1$ particles as $k^{(1)}_j$ and that of the $v_2$ particles as $k^{(2)}_j$, where $j$ numbers the particles. Then, following the theory of integrable systems~\cite{Korepin1993}, we can construct the most general eigenstate of the Hamiltonian Eq.~\eqref{q3wi},
\begin{eqnarray}  \label{eqn:12}
\left| \Psi \right>_{N_1, N_2} & = &  \int \prod_{i=1}^{N_1} dx_i  \prod_{i=1}^{N_2} dy_i \, \psi(x_1, \dots x_{N_1}; y_1, \dots, y_{N_2}) \cr && \times
\hat a^\dagger_1(x_1) \dots \hat a^\dagger_1(x_{N_1}) \, \hat a^\dagger_2(y_1) \dots \hat a^\dagger_2 (y_{N_2}) \left|0 \right> \cr && + \dots,
\end{eqnarray} where $\psi$ is given by
		\begin{align}
		\psi(x_1,\dots x_{N_1}&;y_1\dots y_{N_2})=\mathcal{A}\prod_{j=1}^{N_1}e^{i k^{(1)}_j x_j}\,\prod_{j=1}^{N_2}e^{i k^{(2)}_j y_j} \nonumber \\
		&\times \prod_{m,n} \left[\theta(y_m - x_n) + \theta(x_n - y_m) S(k^{(1)}_n,k^{(2)}_m) \right]
		\end{align}
where $\mathcal{A}$ denotes anti-symmetrization over all $x_j$ and over all $y_j$. The dots at the end of Eq.~\eqref{eqn:12} stand for terms where pairs of fermions are replaced by bosons, in analogy with Eq.~\eqref{eq:ne2}.

It is straightforward to check that Eq.~\eqref{eqn:12} is indeed an eigenstate with energy
        \begin{equation}
        E = v_1 \sum_{j = 1}^{N_1} k^{(1)}_j + v_2 \sum_{j_1}^{N_2} k^{(2)}_j .
        \end{equation}
Crucial in this verification is the fact that the Q3WI Hamiltonian only includes first derivatives of the annihilation operators, unlike standard Hamiltonians, such as Eq.~\eqref{eqn:01}, which
involve second derivatives of the annihilation operators. As a result, the three particle terms which typically need to cancel for a one-dimensional problem to be integrable are absent in our problem.
        
Imposing periodic boundary conditions, with $L$ the system size, we deduce the Bethe-Ansatz equations
		\begin{align}
		\label{eqn:14}
		e^{i k^{(1)}_j L} \prod_{m = 1}^{N_2} \frac{k^{(1)}_j(v_1 - u) + k^{(2)}_m(v_2 - u) - \nu - i\frac{g^2}{2(v_1 - v_2)}}{k^{(1)}_j(v_1 - u) + k^{(2)}_m(v_2 - u) - \nu + i\frac{g^2}{2(v_1 - v_2)}} & = 1 ,\nonumber \\
		e^{i k^{(2)}_j L} \prod_{m = 1}^{N_1} \frac{k^{(1)}_m(v_1 - u) + k^{(2)}_j(v_2 - u) - \nu + i\frac{g^2}{2(v_1 - v_2)}}{k^{(1)}_m(v_1 - u) + k^{(2)}_j(v_2 - u) - \nu - i\frac{g^2}{2(v_1 - v_2)}} & = 1 .
		\end{align}
Taking the logarithm, we find
		\begin{align}
		\label{eqn:15}
		k^{(1)}_j & =  \frac{2 \pi}{L} I_j - \frac{2}{L}\sum_{m=1}^{N_2} \tan^{-1}\left( \frac{k^{(1)}_j(v_1 - u) + k^{(2)}_m(v_2 - u) - \nu}{\frac{g^2}{2(v_1 - v_2)}}\right) ,\nonumber \\
		k^{(2)}_j & =  \frac{2 \pi}{L} J_j + \frac{2}{L}\sum_{m=1}^{N_1} \tan^{-1}\left( \frac{k^{(1)}_m(v_1 - u) + k^{(2)}_j(v_2 - u) - \nu}{\frac{g^2}{2(v_1 - v_2)}}\right) ,
		\end{align}
where the $I_j$ ($J_j$) are distinct half-integers if $N_1$ ($N_2$) is odd, and distinct integers if $N_1$ ($N_2$) is even. We can thus parametrize the solutions to the Bethe equations by sets of integers $I_j$ and $J_j$, which act as quantum numbers for the momenta, $k^{(1)}$ and $k^{(2)}$ respectively. The existence of a unique solution is proved in Ref.~\cite{Williams1998}, where it is demonstrated that the action corresponding to these equations is bounded below and always has a unique minimum in the thermodynamic limit, which we now proceed to study.


		\section{Thermodynamics and Excitation Spectrum}
		\label{TBA}
		
We are now interested in taking the thermodynamic limit of the Bethe Ansatz equations derived in the previous section. To this end, we promote $I_j$ and $J_j$ to continuous variables $I$ and $J$, with $k^{(1)}$ and $k^{(2)}$ now becoming functions of $I$ and $J$ respectively, 
rather than being labelled by a subscript indicating a particle number. Henceforth, it will be convenient to change notations $k^{(1)} \rightarrow k_1$, $k^{(2)} \rightarrow k_2$, 
with $k_1$ and $k_2$ simply denoting the momenta of each species of fermionic particles. 

We then differentiate the Bethe Ansatz equations with respect to the momenta $k_1$ and $k_2$, and introduce the density of vacancies (where a full vacancy is a particle and an empty one is a hole) for the $k_1$ and $k_2$ particles
		\begin{equation}
		\rho^t_1(k_1) = \frac{1}{L} \frac{dI}{d k_1} ,\quad \rho_2^t(k_2) = \frac{1}{L} \frac{dJ}{d k_2},
		\end{equation}
where $\rho_1^t(k_1) = \rho_1(k_1) + \rho^h_1(k_1)$ and $\rho^t_2(k_2) = \rho_2(k_2) + \rho_2^h(k_2)$, with $\rho_{1,2}$ and $\rho^h_{1,2}$ denoting the densities of full and empty vacancies of each particle species respectively. This leads to the following set of coupled integral equations for the densities
		\begin{align}
		\label{eqn:17}
		\rho_1^t(k_1) & =  \frac{1}{2\pi} - \frac{u - v_1}{2\pi}\int dk_2\,\rho_2(k_2)K(k_1,k_2) ,\nonumber \\
		\rho_2^t(k_2) & =  \frac{1}{2\pi} + \frac{u - v_2}{2\pi}\int dk_1 \,\rho_1(k_1)K(k_1,k_2),
		\end{align}
where we have defined the integral kernel,
        \begin{equation}
        \label{eqn:18}
        K(k_1, k_2) = \frac{4 g^2 (v_1 - v_2)}{g^4 + 4 (v_1 - v_2)^2 \left(k_1(u - v_1) + k_2(u - v_2) + \nu\right)^2}.
        \end{equation}
Since the energy of the fermions is unbounded below owing to their chiral nature, we have to implement momentum cut-offs. If, for instance, $v_1, v_2$ are both positive, these cut-offs can be implemented in the following way,
		\begin{align}
		\rho_1^t(k_1) & =  \frac{1}{2\pi} - \frac{u - v_1}{2\pi}\int_{-\Lambda}^{\infty} dk_2\,\rho_2(k_2) K(k_1,k_2), \nonumber \\
		\rho_2^t(k_2) & =  \frac{1}{2\pi} + \frac{u - v_2}{2\pi}\int_{-\Lambda}^{\infty} dk_1\,\rho_1(k_1) K(k_1, k_2).
		\end{align}
Here, the cut-off $\Lambda$ removes particles with large negative kinetic energies. If $v_1$ is negative, we instead cut-off large positive momenta $k_1$, and similarly, if $v_2 < 0$, we cut-off large positive momenta $k_2$. 

To determine the ground state and excitation spectrum of the system, we follow the method introduced by Yang and Yang~\cite{Yang1969}, where we first find the excitation spectrum at finite temperature and then take the limit $T\to 0$ (see~\cite{Korepin1993,Takahashi1999} for a review).

The total number of particles of type $v_1$ and $v_2$ can be found by integrating the density functions
        \begin{equation}
        \label{pdensity}
        N_1/L =  \int d k_1\, \rho_1(k_1),\quad N_2/L =  \int d k_2\, \rho_2(k_2),
        \end{equation}
while the total energy of the system is given by
        \begin{equation}
        \label{totenergy}
        E/L  = v_1\int dk_1\,\rho_1(k_1)\,k_1 + v_2\int dk_2\,\rho(k_2)\, k_2.
        \end{equation}
The entropy of the system is
        \begin{multline}
        S/L =  \int dk_1\,\left[\rho_1^t \log(\rho_1^t) - \rho_1\log(\rho_1) - \rho_1^h\log(\rho_1^h) \right] + \\ \int dk_2\,\left[\rho_2^t \log(\rho_2^t) - \rho_2\log(\rho_2) - \rho_2^h\log(\rho_2^h) \right].
        \end{multline}
We introduce the thermodynamic potential $\Omega$ = $E - T S - h_1 N_1 - h_2 N_2$, where $T$ is the temperature, and $h_1$ and $h_2$ are the chemical potentials conjugate to the conserved particle numbers $N_1$ and $N_2$, defined in Eq.~\eqref{eq:n1n2}. We then minimize $\Omega$ with respect to $\rho_1$ and $\rho_2$, recalling that
		\begin{align}
		\delta\rho_1^t(k_1) &=  -\frac{u - v_1}{2\pi}\int dk_2\,K(k_1,k_2) \delta\rho_2(k_2), \nonumber \\
		\delta\rho_2^t(k_2) &=   \frac{u - v_2}{2\pi}\int dk_1\,K(k_1, k_2) \delta\rho_1(k_1),
		\end{align}
and find a system of coupled non-linear integral equations, usually called the Thermodynamic Bethe Ansatz (TBA) equations,
		\begin{align}
		\label{eqn:24}
		\epsilon_1(k_1) & =  k_1 v_1 - h_1 \nonumber \\ 
		& - T \frac{u - v_2}{2\pi}\int d k_2\,K(k_1, k_2) \ln(1 + e^{-\frac{\epsilon_{2}(k_2)}{T}}), \nonumber \\
		\epsilon_2(k_2) & =  k_2 v_2 - h_2 \nonumber \\
		& + T \frac{u - v_1}{2\pi} \int d k_1\,K(k_1, k_2) \ln(1 + e^{-\frac{\epsilon_{1}(k_1)}{T}}).
		\end{align}
Here, following~\cite{Korepin1993} we have defined
		\begin{equation}
		\label{eqn:25}
		\frac{\rho_1^h (k_1)}{\rho_1 (k_1)} = e^{\frac{\epsilon_{1}(k_1)}{T}},\quad \frac{\rho_2^h (k_2)}{\rho_2 (k_2)} = e^{\frac{\epsilon_{2}(k_2)}{T}},
		\end{equation}
with $\epsilon_1 (k_1 )$ and $\epsilon_2(k_2)$ playing the role of the excitation spectrum of the $v_1$ and $v_2$ particles respectively. In principle, this set of equations produces the excitation spectrum of the system, but being non-linear, there exists no analytic solution. 

Since we are interested in the zero temperature limit here, taking $T\to0$ we find that
		\begin{align}
		\label{eqn:26}
		\epsilon_1(k_1) & =  k_1 v_1 - h_1 + \frac{u - v_2}{2\pi} \int_{\epsilon_2(k_2)<0} dk_2 \,K(k_1, k_2) \, \epsilon_2( k_2), \nonumber \\
		\epsilon_2(k_2) & =  k_2 v_2 - h_2 - \frac{u - v_1}{2\pi} \int_{\epsilon_1(k_1)<0} d k_1\,K(k_1,k_2) \,  \epsilon_1(k_1).
		\end{align}
Here, the integrals are restricted over the regions where $\epsilon_1(k_1)$, $\epsilon_2(k_2) <0$. In addition, from Eq.~\eqref{eqn:25} we see that in the zero temperature limit, 
		\begin{align}
		\epsilon < 0 &\implies \rho^h = 0 ,\nonumber \\
		\epsilon > 0 &\implies \rho = 0 ,
		\end{align}
with the Fermi-momenta, $k_1^f, k_2^f$, defined in such a way that
		\begin{equation}
		\epsilon_1(k_1 = k_1^f) = 0 ,\quad \epsilon_2(k_2 = k_2^f) = 0.
		\end{equation}
From these definitions, we see that if $\epsilon(k) > 0$, then the excitation corresponds to a particle; otherwise, it corresponds to a hole with energy $-\epsilon(k)$. 

For an arbitrary interaction strength $g$, these equations can only be solved numerically, but in what follows we develop analytic solutions for the limits of infinite and weak interactions at zero temperature. 

		
		\section{Strongly Interacting Regime}
		\label{strongint}
		The strongly interacting regime of the Q3WI model corresponds to the broad resonance regime and is accessed by taking the limit $g\, ,\nu\to\infty$, while keeping the ratio $g^2/\nu = \gamma$ fixed. 
In this limit, we can integrate out the bosonic molecules $b$, and the Q3WI Hamiltonian (Eq.~\eqref{q3wi}) reduces to
		\begin{equation} 
		\label{eq:strint} 
	\hat{H} = \int dx  \left[ \sum\limits_{\sigma=1,2}  \hat{a}_\sigma^\dagger \left(-i v_\sigma \partial_x \right) \hat{a}_\sigma -\gamma \hat{a}_1^{\dagger}\hat{a}_2^{\dagger}\hat{a}_2\hat{a}_1 \right].
		\end{equation}
We note that in the conventional case with quadratic disperson Eq.~\eqref{eq:ocm}, this limit is conventionally referred to as the limit of broad resonance, with the term ``strong interactions" reserved for the case when 	
$\left| \gamma \right|$ is large. We will see however that in our case, where particles disperse linearly, the physics of Eq.~\eqref{eq:strint} does not depend much on the magnitude of $\gamma$; we thus refer to Eq.~\eqref{eq:strint}  with arbitrary $\gamma$ as the strongly interacting regime.
		
Hence, in the strongly interacting regime, the Q3WI model describes two species of chiral fermions interacting through a short ranged $\delta$-function interaction, whose strength $\gamma$ is controlled by the parameters $g,\nu$ of the original resonant system. Given the simplicity of this limit, we first study the model Eq.~\eqref{eq:strint} directly and then check that the Bethe Ansatz solution developed in the previous section indeed reproduces this limit correctly. 
		
In order to study the scattering of two different species of fermions (since Fermi statistics prevent two particles of the same species from interacting), we consider the two-particle wave function
		\begin{equation}
		\ket{\Psi}_2 = \int dx\,dy \, \psi(x,y) \, \hat a_1^\dagger(x) \hat a_2^\dagger(y)\ket{0}.
		\end{equation}
This leads to the Schr\"odinger equation
		\begin{equation}
		\label{eqn:31}
		\left(-i v_1 \frac{\partial}{\partial x} - i v_2 \frac{\partial}{\partial y} \right) \psi(x,y) - \gamma  \, \delta(x - y) \psi(x,y)= E \psi(x,y).
		\end{equation}
Since $\hat{a}_1^\dagger$ and $\hat{a}_2^\dagger$ describe distinguishable species of chiral fermions, we should not expect the solution to obey any particular symmetries. Indeed, the solution to Eq.~\eqref{eqn:31} is
		\begin{equation}
		\psi(x,y) = e^{i k_1 x + i k_2 y} \left(\theta(y-x) + S \theta(x - y) \right),
		\end{equation}
where the energy $E = k_1 v_1 + k_2 v_2$. Integrating Eq.~\eqref{eqn:31} over the range $x = y \pm \epsilon$, we find that
		\begin{equation}
		\lim_{\epsilon\to0}\left(-i v_1 + i v_2\right) \psi(x,y)|_{x=y-\epsilon}^{x=y+\epsilon} = \gamma \psi(x,y)|_{x=y},
		\end{equation}
which leads to the $S$-matrix
		\begin{equation}
		\label{eqn:34}
		S = \frac{2(v_1 - v_2) + i\gamma}{2(v_1 - v_2) - i \gamma}.
		\end{equation}
Interestingly, this $S$-matrix is completely independent of momenta. Since $S = e^{i \theta}$ is described by a constant phase shift, it demonstrates that the only effect of the short range $\delta$-function potential on the scattering of chiral fermions is to shift the phase of the outgoing wave relative to the incoming wave, regardless of whether the interaction is repulsive or attractive. 

Another interesting feature of this solution is that 
both the limit $\gamma \rightarrow 0$ and $\gamma\rightarrow \infty$ describe two species of non-interacting fermions. In particular,
		\begin{equation}
		\lim_{\gamma\to0} S = 1,
		\end{equation}
so that the solution to Eq.~\eqref{eqn:31} becomes
		\begin{equation}
		\psi(x,y) = e^{i k_1 x + i k_2 y}.
		\end{equation}
Since $\psi(x,x)\neq 0$ and since the particles are allowed to occupy the same momentum state, the limit $\gamma \to 0$ clearly describes two distinct species of non-interacting fermions. Similarly, in the opposite limit
		\begin{equation}
		\lim_{\gamma \to \pm \infty} S = -1,
		\end{equation}
so that Eq.~\eqref{eqn:31} is solved by 
		\begin{equation}
		\psi(x,y) = e^{i k_1 x + i k_2 y}(\theta(y-x) - \theta(x-y)).
		\end{equation}
In this case, $\psi(x,x) = 0$ but the particles are still allowed to occupy the same momentum state. Thus, in this regime, regardless of whether the interaction is attractive or repulsive, Eq.~\eqref{eq:strint} describes two non-interacting Fermi-gases. This is reflected in the fact that the $S$-matrix is momentum-independent. 

Since we have at our disposal the Bethe ansatz solutions Eq.~\eqref{eqn:14} to the full Q3WI model Eq.~\eqref{q3wi}, it is instructive to verify that they indeed reproduce the broad resonance regime discussed above. First, we observe that in the limit $g^2/\nu\to\gamma$, Eq.~\eqref{eqn:11} reduces to Eq.~\eqref{eqn:34}, reproducing the constant $S$-matrix as expected. We thus find that the Bethe ansatz equations reduce to
		\begin{align}
		k^{(1)}_j & =  \frac{2 \pi}{L} I_j + \frac{2 N_2}{L} \tan^{-1}\left( \frac{2(v_1 - v_2)}{\gamma}\right), \nonumber \\
		k^{(2)}_j & =  \frac{2 \pi}{L} J_j - \frac{2 N_1}{L} \tan^{-1}\left( \frac{2(v_1 - v_2)}{\gamma}\right),
		\end{align}
from which we deduce the densities of the chiral fermions in the broad resonance regime,
		\begin{equation}
		\rho_1^t(k_1) = \rho_2^t(k_2) = \frac{1}{2\pi}, \quad \forall \gamma.
		\end{equation}
Similarly, if we now consider Eqs.~\eqref{eqn:24} in this limit, the integral kernel vanishes, leading to the excitation spectra
		\begin{align}
		\epsilon_1(k_1) & =  k_1 v_1 - h_1 , \nonumber \\
		\epsilon_2(k_2) & =  k_2 v_2 -  h_2 ,
		\end{align}
from which we can deduce the finite temperature densities of the fermions,
		\begin{align}
		\rho_1(k_1) & =  \frac{1}{2\pi}\,\frac{1}{e^{\frac{k_1 v_1 - h_1}{T}} + 1},	\nonumber \\
		\rho_2(k_2) & =  \frac{1}{2\pi}\,\frac{1}{e^{\frac{k_2 v_2 - h_2}{T}} + 1}.	
		\end{align}
We recognize in these two independent Fermi-Dirac distributions.

Thus, unlike the usual case of non-chiral fermions with quadratic dispersion, where at infinite repulsion the fermions behave as a single non-interacting species~\cite{Takahashi1999}, here we find that the strongly interacting (or broad resonance) regime corresponds to both species of fermions surviving and becoming completely non-interacting. In particular, the short ranged $\delta$-function interaction between the two species amounts to a constant scattering amplitude, which leaves the dispersion completely unchanged, with no renormalization of their velocities. 


		\section{Weakly Interacting Regime}
		\label{weakint}

The weakly interacting limit of the Q3WI model Eq.~\eqref{q3wi} corresponds to the limit $g \to 0$. The system is described by the Bethe Ansatz solutions developed in Sec.~\ref{TBA}, where the integral kernel~\eqref{eqn:18} in this limit reduces to  
        \begin{align}
        \label{eqn:43}
        K(k_1,k_2) & = 2 \pi \text{ sgn}(v_1 - v_2)\,\delta\left(k_1(u - v_1) + k_2(u - v_2) + \nu\right) \nonumber \\ &+ \frac{g^2}{v_1 - v_2}\frac{1}{\left(k_1(u - v_1) + k_2(u - v_2) + \nu\right)^2} + O(g^4).
        \end{align}
The leading order ($g$-independent) term of the integral kernel clearly depends on the relative velocities of the chiral fermions $v_1$ and $v_2$, but it also depends on the relative strength of the bosonic velocity $u$ through the $\delta$-function. Owing to this, we find that there are two distinct regimes that must be separately analyzed 
		\begin{enumerate}[{A.}]
    	\item $u > v_1 > v_2$,
    	\item $v_1 > u > v_2$.
		\end{enumerate}
Here, we assume without loss of generality that $v_1>v_2$ since the cases where $v_2>v_1$ simply involve an exchange of fermions of type $1$ and $2$. Furthermore, the parity $x \rightarrow -x$ transformation changes the sign of all three velocities $u$, $v_1$, and $v_2$, so that cases such as $v_1>v_2>u$ can be obtained from case A by a combination of parity and fermion type exchange.

For each of these cases, we are interested in understanding the ground state properties and excitation spectra of the chiral fermions in the zero temperature limit. We consider case A in Sec.~\ref{linear}. We find that the leading order behavior of the excitation spectra Eq.~\eqref{eqn:24} at finite $T$ can be established by considering only the $\delta$-function term in the integral kernel $K(k_1,k_2)$. From these, we can establish a finite $T$ solution for the densities, with the ground state densities found by taking the limit $T \to 0$ (despite what one might na\"ively assume, solving the Bethe Ansatz equations in this case directly at $T=0$ is more difficult than working at $T>0$ and then taking the limit $T \rightarrow 0$). We find that the excitation spectra of the chiral fermions are piece-wise linear functions of momenta. Our procedure then allows us to identify and describe several phases and phase transitions in our system.

For case B, considered in Sec.~\ref{condensate}, where the molecular velocity is bracketed by that of the atoms, we find that simply considering the $\delta$-function term is insufficient and that we must also account for the O($g^2$) term to establish the leading order solution (which is why the $O(g^2)$ term was retained in Eq.~\eqref{eqn:43}). In this case, in order to find the excitation spectra analytically, we make an ansatz that the excitation spectra change sign only once. This ansatz gives us a consistent solution only in the cases where $v_1$ and $v_2$ have the same sign. The solution to the Bethe Ansatz equations found via our procedure is reminiscent of the solution to similar equations that describe the weakly interacting Bose gas, which is in a regime resembling Bose condensation. In particular, the weakly interacting Bose gas can be described with good accuracy by the classical one-dimensional Gross-Pitaevskii (non-linear Schr\"odinger) equation despite the formal absence of Bose condensation in one-dimensional space. Since our solution for the Q3WI model in case B closely resembles this behavior, we term the phase in case B the ``quasi-condensate" phase of our model.

For the case where the atomic velocities have opposite signs while the molecular velocity lies between them, we were not able to find an analytic solution since our ansatz leads to a physically inconsistent solution. 


        \section{Linear, gapless behavior at weak interactions (case A)}
        \label{linear}
        
We first study the cases where the molecular velocity is greater than that of the chiral fermions. Specifically, we consider case A, which corresponds to $u>v_1>v_2$. In order to study this regime, we first define the reduced integral kernel
    \begin{eqnarray}
    \label{reducedK}
	\tilde{K}(k_1,k_2) &=& \lim_{g \rightarrow 0} K(k_1,k_2) \cr &=& 2\pi\,\delta\left(k_1(u - v_1) + k_2(u - v_2) + \nu \right),
	\end{eqnarray} 
since it is sufficient to replace $K$ by its limiting $g \rightarrow 0$ value to find the ground state properties of the Q3WI model at weak interaction strengths within this regime. The TBA equations~\eqref{eqn:25} are thus
     	\begin{align}
		\label{eqn:44}
		\epsilon_1(k_1) & =  k_1 v_1 - h_1 \nonumber \\ 
		&- T \frac{(u - v_2)}{2\pi} \int d k_2\,\tilde{K}(k_1, k_2) \log(1 + e^{-\frac{\epsilon_2(k_2)}{T}}), \nonumber \\
		\epsilon_2(k_2) & =  k_2 v_2 - h_2 \nonumber \\ 
		&+ T \frac{(u - v_1)}{2\pi} \int d k_1\,\tilde{K}(k_1, k_2) \log(1 + e^{-\frac{\epsilon_1(k_1)}{T}}),
		\end{align}
while the densities satisfy the coupled integral equations
       	\begin{align}
		\label{eqn:45}
		\rho^t_1(k_1) & =  \frac{1}{2\pi} - \frac{(u - v_1)}{2\pi}\int dk_2 \,\rho_2(k_2)\tilde{K}(k_1, k_2), \nonumber \\
		\rho^t_2(k_2) & =  \frac{1}{2\pi} + \frac{(u - v_2)}{2\pi}\int dk_1 \,\rho_1(k_1)\tilde{K}(k_1, k_2).
		\end{align} 
We have to supplement the integrals with a momentum cut-off $\Lambda \gg 1$, since the energies are unbounded from below (above) if $v_i$ is positive (negative). Since we are working with the reduced kernel $\tilde{K}$, Eqs.~\eqref{eqn:44} simplify,
        \begin{align}  
        \epsilon_1(k_1) &= k_1 v_1 - h_1 \nonumber \\
        &- T \log\left(1 + \exp\left( -\frac{ \epsilon_2\left(-\frac{k_1 (u - v_1) + \nu}{u - v_2}\right)}{T} \right)\right), \nonumber \\
        \epsilon_2(k_2) &= k_2 v_2 - h_2 \nonumber \\
        &+ T \log\left(1 + \exp\left( -\frac{ \epsilon_1\left(-\frac{k_2 (u - v_2) + \nu}{u - v_1}\right)}{T} \right)\right).
        \end{align}
This is valid for arbitrary $k_1$ and $k_2$. In order to simplify these equations further, we set 
\begin{equation}
\tilde{k}_2 = -\frac{k_1(u-v_1)+\nu}{u-v_2}, 
\end{equation}
and find that
\begin{align}
        \epsilon_1(k_1) &= k_1 v_1 - h_1 - T \log\left(1 + \exp\left( -\frac{ \epsilon_2\left(\tilde{k}_2  \right) }{T}\right)\right), \nonumber \\
        \epsilon_2(\tilde{k}_2) &= \tilde{k}_2 v_2 - h_2 + T \log\left(1 + \exp\left( -\frac{ \epsilon_1\left(k_1 \right)}{T} \right)\right).
        \end{align}
Equivalently, 
        \begin{equation}
        e^{-\frac{\epsilon_1(k_1)}{T}} = e^{- \frac{k_1 v_1 - h_1}{T}} \left(1 + \frac{\exp\left(\frac{\frac{k_1(u - v_1) + \nu}{u - v_2} v_2 + h_2}{T} \right)}{1 + e^{-\frac{\epsilon_1(k_1)}{T}}} \right).
        \end{equation}
Solving this quadratic equation for $\exp \left(-\epsilon_1(k_1)/T \right)$ leads to
       \begin{widetext}
        \begin{equation}
        \label{eqn:47}
        \exp\left(-\frac{\epsilon_1(k_1)}{T}\right) = \frac{1}{2}\left(\exp\left(-\frac{k_1 v_1 - h_1}{T}\right) - 1\right) + \frac{1}{2} \sqrt{\left(\exp\left(-\frac{k_1 v_1 - h_1}{T}\right)  + 1 \right)^2 + 4\, \exp\left( -\frac{k_1 v_1 - \frac{k_1 (u - v_1) + \nu}{u - v_2} v_2 - h_1 - h_2}{T} \right)}.
        \end{equation}
        \end{widetext}
Here, we must choose the positive sign in front of the square root in order to keep the whole expression positive. This gives us the finite temperature solution for the excitation spectrum $\epsilon_1(k_1)$ of the first species of chiral fermions. The spectrum $\epsilon_2(k_2)$ of the second species of fermions can then be found by substituting Eq.~\eqref{eqn:47} back into the equation for $\epsilon_2(k_2)$. 

Since we have established the finite temperature solution for the spectra, we can also find the solutions for the densities from Eq.~\eqref{eqn:45}. Recalling that $\rho_i^t = \rho_i + \rho_i^h$ and using Eq.~\eqref{eqn:25}, these simplify to
        \begin{align}
        \rho_1(k_1) \left(1 + e^{\frac{\epsilon_1(k_1)}{T}} \right) &= \frac{1}{2\pi} - \frac{u - v_1}{u - v_2} \rho_2 \left(- \frac{k_1(u - v_1) + \nu}{u - v_2}\right), \nonumber \\
        \rho_2(k_2) \left(1 + e^{\frac{\epsilon_2(k_2)}{T}} \right) &= \frac{1}{2\pi} + \frac{u - v_2}{u - v_1} \rho_1 \left(- \frac{k_2(u - v_2) + \nu}{u - v_1}\right).
        \end{align}
      	These are a set of simple algebraic equations which can be solved in order to find that 
        \begin{align}
        \label{density}
        &\rho_1(k_1) = \nonumber \\
        &\frac{\frac{1}{2\pi}\left(1 + \exp\left(\frac{\epsilon_2\left(-\frac{k_1 (u - v_1) + \nu}{u - v_2}\right)}{T} \right) \right)- \frac{1}{2\pi}\frac{u - v_1}{u - v_2}}{1 + \left(1 + \exp\left(\frac{\epsilon_2\left(-\frac{k_1 (u - v_1) + \nu}{u - v_2}\right)}{T} \right) \right)\left(1 + \exp\left(\frac{\epsilon_1(k_1)}{T} \right) \right) }, \nonumber \\
\nonumber \\
        &\rho_2(k_2) = \nonumber \\
        &\frac{\frac{1}{2\pi}\left(1 + \exp\left(\frac{\epsilon_1\left(-\frac{k_2 (u - v_2) + \nu}{u - v_1} \right)}{T} \right) \right) + \frac{1}{2\pi}\frac{u - v_2}{u - v_1}}{1 + \left(1 + \exp\left(\frac{\epsilon_1\left(-\frac{k_2(u-v_2)+\nu}{u-v_1}\right)}{T} \right)\right)\left(1 + \exp\left(\frac{\epsilon_2(k_2)}{T} \right)\right)}.
        \end{align}
Equations~\eqref{eqn:47}, \eqref{density} are the solutions to the Bethe Ansatz equations at finite $T$ in case A, studied in this section.

In order to uncover the physics described by these equations, we now examine the $T \rightarrow 0$ limit of the expressions we have obtained thus far. Technically, this limit is simple to take. It follows from observing that $\exp(c/T) \gg 1$ in this limit if $c>0$ and $\exp(c/T) \ll 1$ if $c<0$. The procedure then reduces to evaluating the signs of various combinations of parameters; upon doing this, the problem splits into various cases distinguished by different values of these signs. 

Carrying out this procedure, we find that there are five further cases within case A, $u>v_1>v_2$, that must be considered separately,
\begin{enumerate}
\item $u>v_1>v_2>0$ ,
\item $u>v_1>0>v_2$ :
\begin{itemize}
\item[] 2i) \,$u < \frac{2 v_1 v_2}{v_1+v_2}$,
\item[] 2ii) $u > \frac{2 v_1 v_2}{v_1+v_2}$,
\end{itemize}
\item $u>0>v_1>v_2$,
\item $0>u>v_1>v_2$.
\end{enumerate}

We now discuss the general procedure through which the ground state properties of these systems can be derived, using as an example the case A1 where $u>v_1>v_2>0$. In particular, we will establish the presence of a quantum phase transition in the Q3WI model through this example and leave the details of the transitions in the other cases to App.~\ref{app:D}, since the same method applies to each of these. 

\subsubsection*{Quantum phase transition in the Q3WI model}

As a first step in establishing the ground state properties of the Q3WI model in the weakly interacting limit, we consider particle velocities such that $u > v_1 > v_2 > 0$ and take the zero temperature limit $T\to 0$ of Eq.~\eqref{eqn:47}. This gives us the excitation spectra and ground state densities of the chiral fermions. For both species of fermions, the excitation spectra turn out to be piece-wise linear functions of momenta, with derivatives $\epsilon_i'(k_i)$ that are discontinuous at certain momenta $k_i^{(j)}$ ($j = a,b,c$).
For future reference, these are defined as
		 \begin{align}
       	 \label{parameters}
       	 k_1^{(a)} &= \frac{\left(h_1-h_2\right) \left(u-v_2\right)-\nu  v_2}{u (v_1+v_2) -2 v_1 v_2}, \nonumber \\
       	 k_1^{(b)} &= \frac{\left(h_1+h_2\right) \left(u-v_2\right)+\nu  v_2}{u \left(v_1-v_2\right)}, \nonumber \\
       	 k_1^{(c)} &= \frac{h_2 \left(v_2-u\right)-\nu  v_2}{v_2 \left(u-v_1\right)}, \nonumber \\
       	 k_2^{(a)} &= -\frac{\left(h_1+h_2\right) \left(u-v_1\right)+\nu  v_1}{u \left(v_1-v_2\right)}, \nonumber \\
         k_2^{(b)} &= \frac{\left(h_2-h_1\right) \left(u-v_1\right)-\nu  v_1}{u \left(v_1+v_2\right)-2 v_1 v_2}, \nonumber \\
         k_2^{(c)} &= \frac{h_1 \left(v_1-u\right)-\nu  v_1}{v_1 \left(u-v_2\right)}.
         \end{align}
Furthermore, the ground state densities $\rho_i(k_i)$ are momentum-independent constants that change their values discontinuously at the above momenta $k_i^{(j)}$.

We further find that the behavior of the excitation spectra and ground state densities differs depending on the sign of the parameter
	\begin{equation}
	\label{alpha}
	\alpha(h_1,h_2,\nu) = h_1 v_2 (u-v_1) + h_2 v_1(u - v_2) + \nu v_1 v_2 ,
	\end{equation}
which is controlled by the chemical potentials $h_i$ and the detuning $\nu$. We note that at $\alpha(h_1,h_2)= 0$, the momenta defined in Eq.~\eqref{parameters} satisfy $k_i^{(a)} =k_i^{(b)}=k_i^{(c)}$.

\begin{figure}[t]
    \includegraphics[width=8cm]{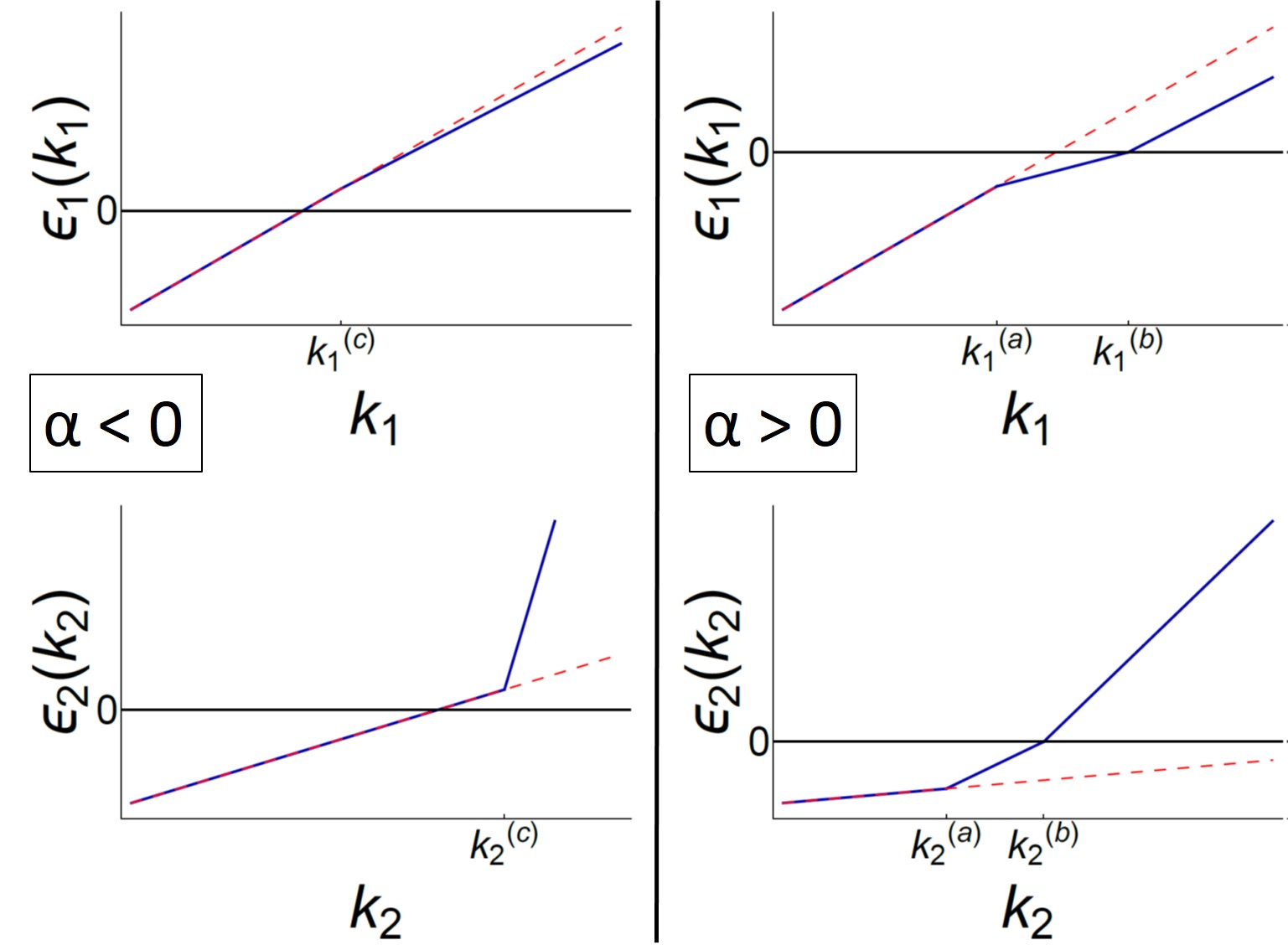}
    \centering
    \caption{\label{fig:01} Excitation spectra for case A1, where the parameters are taken, for the purpose of illustration, as $u = 6,v_1 = 4,v_2 = 1,h_1 = 9,h_2 = 2$. The figures in the left panel correspond to $\alpha<0$ ($\nu = -20$), in which case both species of fermions remain non-interacting throughout the Fermi sea. The figures in the right panel correspond to $\alpha>0$ ($\nu = 20$), and in this case the fermions interact within a finite range of momenta $k_i^{(a)}\leq k_i \leq k_i^{(b)}$. The red dashed lines in either case are the non-interacting excitation spectra, $k_i v_i - h_i$.}
    \end{figure}

For $\alpha \leq 0$, the excitation spectra are
        \begin{align}
        &\epsilon_1(k_1) = \\
        &\begin{dcases}
		k_1 v_1 - h_1, & k_1<k_1^{(c)}, \\
	    \frac{k_1 u (v_1-v_2)-\nu v_2}{u-v_2} - h_1 - h_2, & k_1>k_1^{(c)},
		\end{dcases}\\
		&\epsilon_2(k_2) = \\
		&\begin{dcases}
		k_2 v_2 - h_2, &k_2<k_2^{(c)}, \\
		\frac{k_2 \left( u (v_1+v_2)-2 v_1 v_2 \right) + \nu v_1 }{u-v_1} + h_1-h_2, &k_2>k_2^{(c)}.
		\end{dcases}
        \end{align}
As always, a special role is played by the value of the momenta where the excitation spectra equal zero. We term these the Fermi momenta $k_i^f$. When $\alpha \le 0$, the Fermi momenta are
 \begin{equation}
 k_1^{f} = \frac{h_1}{v_1}, \quad  k_2^{f} = \frac{h_2}{v_2}.
 \end{equation}
 It can be verified that as long as $\alpha<0$,
 \begin{equation} k_i^{c} > k_i^f,
 \end{equation}
 and that $k_i^c=k_i^f$ if $\alpha=0$. The excitation spectra for both species of chiral fermions are shown in Fig.~\ref{fig:01}.

The ground state densities can also be found from the $T\to0$ limit of Eq.~\eqref{density},
        \begin{align}
        \rho_1(k_1) = 
        \begin{dcases}
        \frac{1}{2\pi}, &k_1<k_1^{f}, \\
        0, &k_1>k_1^{f},
        \end{dcases}\\
        \rho(k_2) = 
        \begin{dcases}
        \frac{1}{2\pi}, &k_2<k_2^{f}, \\
        0, &k_2>k_2^{f}.
        \end{dcases}
        \end{align}
        
Let us examine $\epsilon_1(k_1)$ first. We see that in the region $k_1 \leq k_1^{(c)}$, these fermions behave as though they were non-interacting, since their spectrum corresponds to that of non-interacting chiral fermions dispersing with velocity $v_1$. Beyond this range of momenta, the behavior of the fermions changes as the interaction forces them to disperse with a different velocity. Importantly, the Fermi momentum i.e., where $\epsilon_1(k_1^f) = 0$ 
matches the Fermi momentum of a non-interacting system of chiral fermions. Since the behavior of $\epsilon_1(k_1)$ changes at a momentum larger than $k_1^f$, the ground state of the $v_1$ fermions corresponds to a filled Fermi sea of non-interacting chiral fermions, reflected in the ground state density $\rho_1(k_1)$, which equals $1/(2\pi)$, the momentum density for non-interacting free fermions.

Similarly, we observe that the spectrum of the $v_2$ fermions corresponds to that of non-interacting chiral fermions, since $\epsilon_2(k_2) = k_2 v_2 - h_2$ in the momentum range $k_2 \leq k_2^{(c)}$. The ground state of these fermions is thus also described by a filled Fermi sea of non-interacting fermions, with the Fermi momentum $k_2^f$ and ground state density $\rho_2(k_2)$ also reflecting the non-interacting behavior.        

We now consider the regime where $\alpha >0$. Here, we again find that the spectra are piece-wise linear functions
		\begin{align}
		&\epsilon_1(k_1) = \nonumber \\
		&\hspace{-0.3cm}\begin{dcases}
		k_1 v_1 - h_1, & k_1<k_1^{(a)}, \\
		\frac {k_1 u (v_1-v_2) - \nu v_2}{2 (u-v_2)}  - \frac { h_1 + h_2}{2}, &  k_1^{(a)}<k_1<k_1^{(b)}, \\
		\frac {k_1 u (v_1-v_2) - \nu v_2}{ (u-v_2)}  -h_1 - h_2, & k_1>k_1^{(b)}
		\end{dcases} \\
		&\epsilon_2(k_2) = \nonumber \\
		&\hspace{-0.3cm}\begin{dcases}
		k_2 v_2 - h_2, & k_2<k_2^{(a)}, \\
		\frac{k_2 \left( u (v_1+v_2)-2 v_1 v_2 \right) + \nu v_1 }{2(u-v_1)} + \frac{h_1-h_2}{2} , & k_2^{(a)}<k_2<k_2^{(b)} \\
		\frac{k_2 \left( u (v_1+v_2)-2 v_1 v_2 \right) + \nu v_1 }{u-v_1} + h_1-h_2,  & k_2>k_2^{(b)},
		\end{dcases}
		\end{align}
with the ground state densities given by
        \begin{align}
        \rho_1(k_1) = 
        \begin{dcases}
        \frac{1}{2\pi}, &k_1<k_1^{(a)} \\
        \frac{1}{4\pi} \left(1 - \frac{u - v_1}{u - v_2} \right), &k_1^{(a)}<k_1<k_1^{(b)},\\
        0, &k_1>k_1^{(b)},
        \end{dcases} \\
        \rho_2(k_2) = 
        \begin{dcases}
        \frac{1}{2\pi}, &k_2<k_2^{(a)}, \\
        \frac{1}{4\pi} \left(1 + \frac{u - v_2}{u - v_1} \right), &k_2^{(a)}<k_2<k_2^{(b)},\\
        0, &k_2>k_2^{(b)}.
        \end{dcases}
        \end{align}  
 Furthermore, the Fermi momenta, defined by $\epsilon_i(k_i^f)=0$, are now 
 \begin{equation} k_i^f = k_i^{(b)}
 \end{equation}
 as can be easily verified.

We now contrast the behavior of $\epsilon_1(k_1)$ in this regime with that in the previous case. For $\alpha>0$, as depicted in Fig.~\ref{fig:01}, we see that while the system displays non-interacting behavior deep within the Fermi sea i.e., for $k_1 \leq k_1^{(a)}$, this behavior changes before the spectrum crosses zero energy. In this case, the spectrum of the $v_1$ fermions deviates from the non-interacting spectrum in the vicinity of the Fermi momentum.
Importantly, the ground state density $\rho_1(k_1)$ is also modified from the non-interacting value $1/(2\pi)$ in the vicinity of the Fermi momentum $k_1^f$ but retains the non-interacting value for $k_1 \leq k_1^{(a)}$. Hence, the ground state of the system is described by a filled Fermi sea of chiral fermions that, at large negative momenta, behave as non-interacting chiral fermions, but interact close to the Fermi momentum with a modified momentum density and dispersion due to the weak resonant interactions.

Analogous behavior occurs for the $v_2$ particles when $\alpha > 0$, since their spectrum matches that of a non-interacting spectrum for $k_2 \leq k_2^{(a)}$ but the dispersion changes 
in the vicinity of the Fermi momentum $k_2^f$. The ground state density in the vicinity of the Fermi level also deviates from its non-interacting value.

Thus, we have explicitly found the excitation spectrum and the ground state densities for the system. The fact that the ground state properties of these chiral fermions are significantly modified depending on the parameter $\alpha$ is an indication that the system undergoes a quantum phase transition. For $\alpha \leq 0$, the ground state of the system behaves as though it is insensitive to interactions, reflected in the densities $\rho_1,\rho_2$ which remain $1/(2\pi)$, their value for free fermions, all through the Fermi sea. However, for $\alpha > 0$, the solution to our problem changes qualitatively since the particles now interact within a finite range of momenta in the vicinity of $k_i^f$, changing the ground state densities in this region.
   
We can further elucidate the nature of this change if we calculate the particle densities $D_1,D_2$ as functions of the chemical potentials $h_1,h_2$. We can then invert this relation to find the chemical potentials as functions of particle density, i.e., $h_i(D_1,D_2)$ and look for discontinuities in these functions or their derivatives. 

The particle densities are defined as
        \begin{equation}
        D_1 = \int_{-\Lambda}^{\infty} dk_1 \,\rho_1(k_1)- D_1^{(0)}, \quad D_2 = \int_{-\Lambda}^{\infty} d k_2\,\rho_2(k_2) - D_2^{(0)}.
        \end{equation}
Here, $D_i^{(0)}$ are the densities calculated at some reference point to cancel the contribution from the cut-off, $\Lambda$. We can also get the total energy density of the system, 
        \begin{equation}
        \mathcal{E} = v_1 \int_{-\Lambda}^{\infty} d k_1\,k_1 \,\rho_1(k_1) + v_2 \int_{-\Lambda}^{\infty} d k_2\, k_2 \,\rho_2(k_2) - \mathcal{E}_1^{(0)} - \mathcal{E}_2^{(0)},
        \end{equation}
and define the inverse compressibility as
        \begin{equation}
        \kappa_{ij}^{-1} = \frac{\partial^2 \mathcal{E}}{\partial D_i \partial D_j} = \frac{\partial h_i}{\partial D_j}, \quad i,j = 1,2
        \end{equation}
which satisfies $\kappa_{ij}^{-1} = \kappa_{ji}^{-1}$.

Let us consider the case $\alpha\leq 0$ first. Here, we find that,
        \begin{equation}
        D_1 = \frac{h_1}{2\pi v_1}, \quad D_2 = \frac{h_2}{2 \pi v_2},
        \end{equation}
which when inverted leads to
        \begin{equation}
        h_1(D_1,D_2) = 2 \pi v_1 D_1, \quad h_2(D_1,D_2) = 2 \pi v_2 D_2.
        \end{equation}
Thus, we find the inverse compressibility
        \begin{equation}
        \kappa^{-1} = \left( 
        \begin{array}{cc}
         2\pi v_1 & 0 \\
         0 & 2\pi v_2
        \end{array} \right),
        \end{equation}
which corresponds to the compressibility of a system of two independent species of non-interacting chiral fermions. 

Next we consider the regime where $\alpha>0$. Here, we find the chemical potentials
        \begin{align}
        h_1 & =  \frac{2 \pi  D_1 \left(u^2-v_1 v_2\right)+\left(u-v_1\right) \left(2 \pi  D_2 \left(u-v_2\right)+\nu \right)}{2 u-v_1-v_2}, \nonumber \\
		h_2 & = \frac{2 \pi  D_2 \left(u^2-v_1 v_2\right)+\left(u-v_2\right) \left(2 \pi  D_1 \left(u-v_1\right)+\nu \right)}{2 u-v_1-v_2},
		\end{align}       
from which we find the inverse compressibility
        \begin{equation}
        \kappa^{-1} = \frac{2\pi}{2u - v_1 - v_2}\left(
        \begin{array}{cc}
        u^2 - v_1 v_2 & (u - v_1)(u - v_2) \\
        (u - v_1)(u - v_2) & u^2 - v_1 v_2
        \end{array}
        \right),
        \end{equation}
which distinctly differs from the compressibility in the regime $\alpha \geq 0$. Thus, even though the spectra remain gapless for all values of $\alpha$, the compressibilities $\kappa_{11}$ and $\kappa_{22}$ are discontinuous, which indicates a phase transition. Since we now have the chemical potentials as functions of the particle densities, we can now also define a parameter
        \begin{equation}
        \beta(D_1,D_2) = 2\pi D_1 (u - v_1) + 2\pi D_2 (u - v_2) + \nu ,
        \end{equation}
such that $\text{sgn}\left(\alpha(h_1,h_2)\right) = \text{sgn}\left(\beta(D_1,D_2)\right)$. For fixed particle densities, as we vary the detuning $\nu$, the discontinuity hence occurs at $\nu = -2\pi D_1 (u - v_1) - 2\pi D_2 (u - v_2)$.

We have thus shown that the Q3WI model in the weakly interacting regime undergoes a phase transition as a function of the particle densities of two species of particles at the point where $\alpha=\beta=0$. The nature of this phase transition is such that the system goes from being completely non-interacting throughout the Fermi sea ($\beta \leq 0$) to being interacting within a finite range of momenta close to the Fermi momenta ($\beta > 0$). Since the compressibility is sensitive primarily to the ground state density in the vicinity of the Fermi level $k_i^f$ and since the system's behavior near $k_i^f$ changes abruptly as a function of $\beta$, it is intuitively clear that this change should manifest itself as a discontinuity in $\kappa$.

Although we have focused on a particular case here ($u>v_1>v_2>0$), we find similar phenomena in all other cases as long as $u > v_1 > v_2$ or $v_1 > v_2 > u$ (with the cases where $v_2 > v_1$ understood by exchanging the $v_1$ and $v_2$ fermions). Specifically, in each case we find that the spectra remain gapless while the compressibilities $\kappa$ display a discontinuity as a function of the parameter $\alpha$ in response to the change in the system's behavior close to the Fermi level. While the discussion of the ground state properties and associated phase transitions for all other cases is left to App.~\ref{app:D}, we now review those cases where the Q3WI model displays rather unconventional behavior.

We also note that while we will only plot the excitation spectra in what follows, the ground state densities may be simply inferred from these, since $\rho_i(k_i)$ equals $1/(2\pi)$ over the momentum range where $\epsilon_i(k_i)$ coincides with its non-interacting value. Conversely, if the spectrum deviates from $k_i v_i - h_i$ over some range of momenta within the Fermi sea, then the corresponding ground state density no longer equals $1/(2\pi)$ and the $v_i$ fermions are hence interacting over this range. By definition, $\rho_i(k_i)$ vanishes for $k_i > k_i^f$.

\subsubsection*{Interacting behavior deep within the Fermi sea}

\begin{figure}[t]
    \includegraphics[width=8cm]{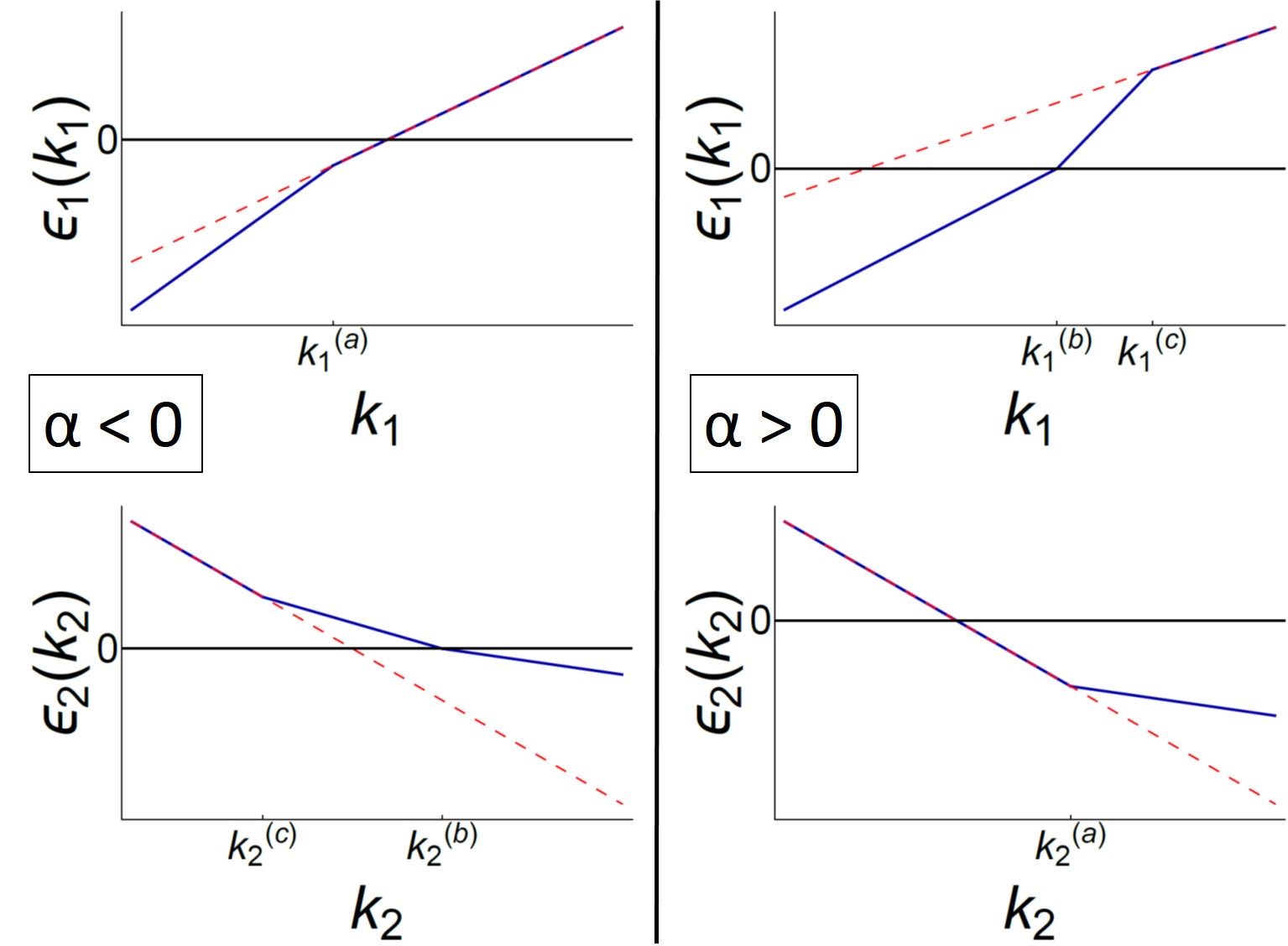}
    \centering
    \caption{\label{fig:02} Excitation spectra for case A2i, with $u = 10,v_1 = 2,v_2 = -10,h_1 = 2,h_2 = 3$. The left-hand panel depicts the regime $\alpha<0$ ($\nu = 20$), where the $v_2$ fermions interact throughout the Fermi sea while the $v_1$ species interacts deep within the Fermi sea but becomes non-interacting for $k_1^{(a)}\leq k_1 \leq k_1^{f}$. The right-hand panel depicts the case $\alpha>0$ ($\nu = -30$), where the behavior is reversed. The red dashed lines are the spectra in the non-interacting case.}
    \end{figure}

In the case considered above, regardless of $\text{sgn}(\alpha)$, both species of chiral fermions displayed non-interacting behavior at large negative momenta, i.e., deep within the Fermi sea. This is in accordance with expectations, since interactions typically only modify the behavior in the vicinity of the Fermi momenta. However, if we change the velocities such that $u > v_1 > 0 > v_2$ with $u(v_1 + v_2) - 2 v_1 v_2 < 0$, i.e., case A2i of our classification, we encounter unusual behavior in the spectrum of both species of fermions, as illustrated in Fig.~\ref{fig:02}.

Specifically, regardless of $\text{sgn}(\alpha)$ we find that both species of fermions remain interacting deep into the Fermi sea since their ground state densities deviate from $1/(2\pi)$ (see App.~\ref{sec:A2a} for details). For $\alpha<0$, we find that the $v_1$ particles display interacting behavior for large negative momenta while remaining non-interacting in the vicinity of $k_1^f$ whereas the $v_2$ particles interact for all momenta $k_2 \geq k_2^f$ (since $v_2 < 0$ here, we impose a positive momentum cut-off on these fermions). That the ground state densities are modified from $1/(2\pi)$ is reflected in the deviation of the fermionic spectra from the non-interacting dispersions. In addition, we find that the densities change at precisely those momenta at which the dispersions change. When $\alpha > 0$, the behavior of the $v_1$ and $v_2$ fermions is reversed since now the $v_1$ particles interact for all $k_1 \leq k_1^f$ while the $v_2$ fermions behave as non-interacting fermions in a finite range of momenta close to $k_2^f$.

The change in behavior is reflected in the compressibilities, which are discontinuous at $\alpha = 0$. For instance, we find that $\kappa_{11}^{-1} = 2 \pi v_1$ for $\alpha < 0$, which is the compressibility associated with non-interacting chiral fermions dispersing with velocity $v_1$. However, for $\alpha > 0$, $\kappa_{11}^{-1}$ changes its value since now these fermions are interacting throughout the Fermi sea. Similar behavior is displayed by the $v_2$ fermions. Thus, in this parameter regime we find that the Q3WI model displays behavior starkly in contrast with that expected from weakly interacting fermions, since in our model the chiral fermions remain interacting far into the Fermi sea. 

\subsubsection*{Interaction induced loss of chirality}

Thus far, both cases encountered have the property that the fermionic spectra remain gapless, and though the velocities with which they disperse change abruptly at certain special momenta, the chirality of each species stays invariant throughout the Fermi sea and matches that of the non-interacting case i.e., $\text{sgn}\left(\epsilon_i'(k_i)\right) = \text{sgn}(v_i)$. If we instead consider the parameter regime where $u>v_1>0>v_2$ with $u(v_1 + v_2) -2 v_1 v_2 > 0$ i.e., case A2ii, then this is no longer the case.
	\begin{figure}[t]
    \includegraphics[width=8cm]{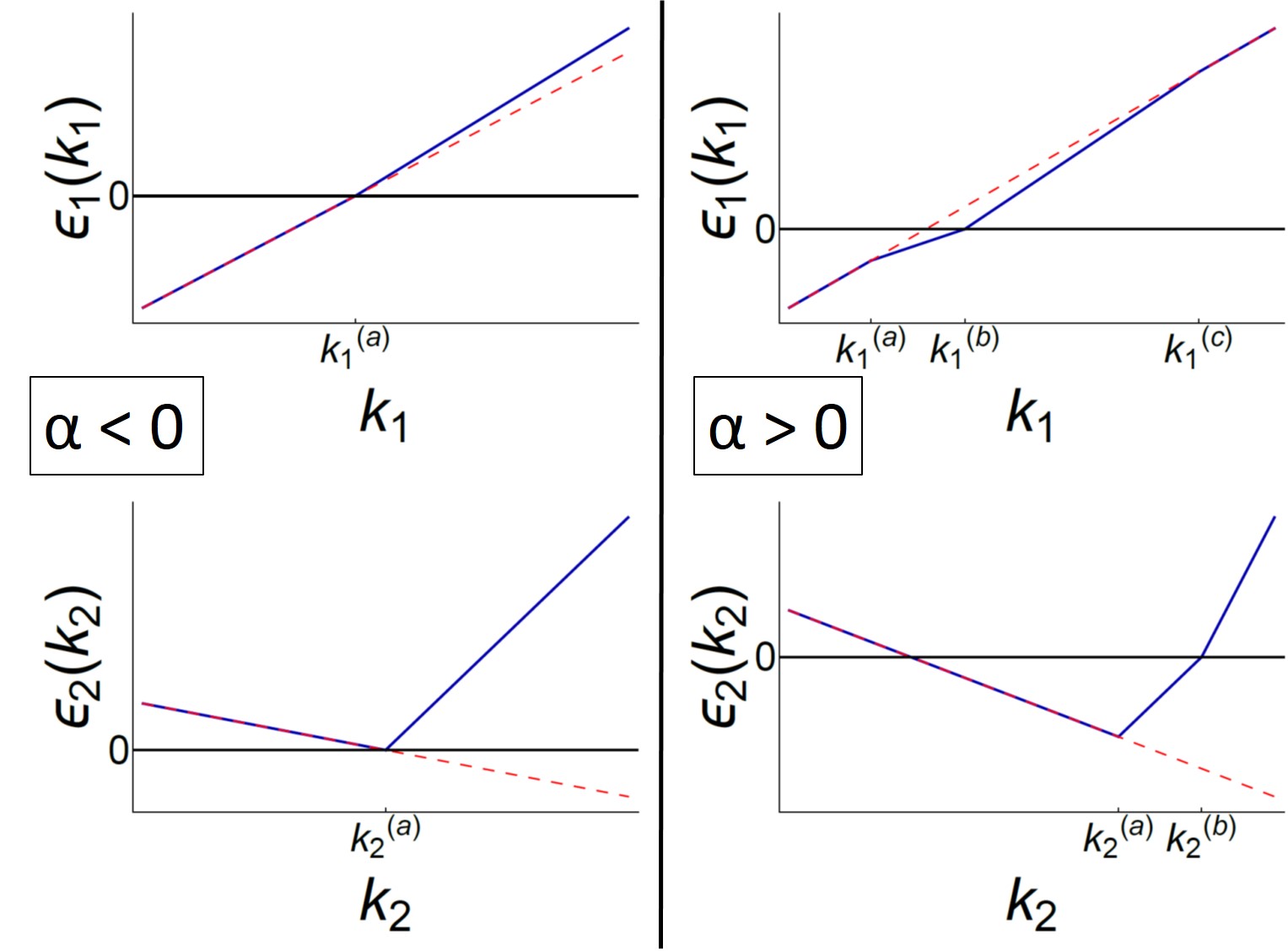}
    \centering
    \caption{\label{fig:03} Excitation spectra for case A2ii, with $u = 6,v_1 = 4,v_2 = -3,h_1 = 2,h_2 = 3$. The plots on the left are for $\nu = 8$, corresponding to $\alpha=0$, while the ones on the right are for $\nu = -10$ corresponding to $\alpha>0$. The red dashed lines are the excitation spectra for a non-interacting system ($g = 0$).}
    \end{figure}
    
For velocities that satisfy the above conditions, we find that the parameter $\beta$ is simply given by the density of the $v_2$ fermions, $\beta = D_2$ (see App.~\ref{sec:A2b} for details). Hence, the case $\beta<0$ (equivalently $\alpha < 0$) is un-physical here and we must instead consider how the system changes when $\beta$ becomes non-zero. Since $\beta = 0$ corresponds to a vanishing density of the $v_2$ particles, we expect that $\kappa_{22}^{-1} = 0$ and that $\kappa_{11}^{-1}$ corresponds to the compressibility of non-interacting chiral fermions, $2 \pi v_1$. Indeed, as shown in Fig.~\ref{fig:03}, $\epsilon_1(k_1)$ coincides with the non-interacting spectrum throughout the Fermi sea and $\rho_1(k_1) = 1/(2\pi)$ in this region. Since $\epsilon_2(k_2) > 0$ for all $k_2$, this implies a vanishing momentum density of the $v_2$ particles, consistent with $D_2 = 0$.

However, as soon as we have a non-zero density $D_2$, the behavior changes drastically. In particular, we see from Fig.~\ref{fig:03} that the $v_1$ particles begin interacting close to the Fermi momentum $k_1^f = k_1^{(b)}$. This change is also reflected in $\kappa_{11}$. More interestingly, we see that the spectrum $\epsilon_2(k_2)$ is no longer chiral and crosses zero energy at two distinct Fermi momenta. The chirality of these fermions changes sign abruptly at $k_2^{(a)}$ and they display non-interacting behavior close to $k_2^{f,1} = h_2/v_2$ but interact in the vicinity of $k_2^{f,2} = k_2^{(b)}$, with the compressibility reflecting this behavior. Thus, we see that in a certain parameter regime, even weak interactions are sufficient to change the chiral nature of the $v_2$ particles. This is remarkable since, typically, we expect that in the absence of back-scattering the velocity should never change sign and yet, we find that this is indeed the behavior exhibited by the Q3WI model in the weakly interacting regime. 

In this section, we have developed a procedure for deriving the weakly interacting, zero temperature ground state densities and excitation spectra of the Q3WI model for case A. In particular, we have demonstrated the existence of a quantum phase transition in this model through an example, with the details of all other cases left to Appendix~\ref{app:D}. Furthermore, we have discussed the qualitatively different behaviors exhibited by this model, focusing on the unconventional nature of the spectra that arise as a consequence of weak interactions. 
      
        \section{Quasi-condensates at weak interactions (case B)}
        \label{condensate}        
        
We now consider the case $v_1 > u > v_2$, where the Q3WI model exhibits qualitatively different behavior compared to that encountered in Sec.~\ref{linear}. While in the previous section we saw that the spectra were piece-wise linear functions of momentum and that the ground state densities were momentum-independent constants, here we find that the dispersions, while still remaining gapless, are no longer linear in momentum and that the densities $\rho_i(k_i)$ are momentum dependent. 

The origin of this change in behavior lies in the fact that when the atomic velocities $v_i$ bracket the molecular velocity $u$, the ground state densities diverge in the limit $g\to 0$. This is analogous to the behavior of the Lieb-Linger model~\cite{LiebLiniger,Lieb1963} in the weakly interacting limit. In our case, we will similarly find that it is insufficient to keep only the leading order $g$-independent $\delta$-function term in the integral kernel $K(k_1,k_2)$~\eqref{eqn:43}. 

In order to show this explicitly, let us consider Eq.~\eqref{eqn:45}, where only the $\delta$-function term in the $g\to 0$ approximation of the kernel (Eq.~\eqref{eqn:43}) is kept in the reduced kernel $\tilde{K}$, given by Eq.~\eqref{reducedK}. We also make the ansatz that the excitation spectrum $\epsilon_i(k_i)$ crosses zero energy only at a single Fermi point $k_i^f$. We will justify our ansatz later for the cases where sgn$(v_1) =$ sgn$(v_2)$; we are unable to find a solution when sgn$(v_1) = -$ sgn$(v_2)$ as this ansatz breaks down.         

In the $T\to0$ limit, the equations for the ground state densities thus become
        \begin{align}
        \label{eqn:124}
        \rho_1(k_1) & = \frac{1}{2\pi} + \frac{v_1 - u }{2 \pi} \int_{-\Lambda}^{k_2^f}dk_2\,\rho_2(k_2)\tilde{K}(k_1,k_2),  \\
        \rho_2(k_2) & = \frac{1}{2\pi} + \frac{u - v_2}{2 \pi} \int_{-\Lambda}^{k_1^f}dk_1\,\rho_1(k_1)\tilde{K}(k_1,k_2), \label{eqn:k12}
        \end{align}
where $\tilde{K}$ is the reduced integral kernel~\eqref{reducedK}. Here, we have assumed without loss of generality that $v_1,v_2>0$ since the case where $v_1,v_2<0$ follows similarly. With the kernel $\tilde{K}$ being a $\delta$-function, the integral in the right hand side of Eq.~\eqref{eqn:124} is non-zero only if
		\begin{equation}
		\frac{-\Lambda(u - v_2) + \nu}{ v_1-u} \leq k_1 \leq \frac{k_2^f(u - v_2) + \nu}{v_1-u} ,
		\end{equation}
while the integral in the right hand side of Eq.~\eqref{eqn:k12} is non-zero only if
		\begin{equation}		
		\frac{-\Lambda ( v_1-u) - \nu}{u - v_2}  \leq k_2 \leq \frac{k_1^f ( v_1 - u) - \nu}{u - v_2}.
		\end{equation}
For $k_i$ that satisfy these conditions, Eq.~\eqref{eqn:124} becomes		
        \begin{align}
        \rho_1(k_1) &= \frac{1}{2\pi} + \frac{v_1 - u}{u - v_2}\,\rho_2\left(\frac{k_1 ( v_1-u)-\nu}{u - v_2}\right), \nonumber \\
        \rho_2(k_2) &= \frac{1}{2\pi} + \frac{u - v_2}{v_1-u}\,\rho_1\left(\frac{k_2(u - v_2) + \nu}{ v_1-u}\right).
        \end{align}
These are a simple set of algebraic equations that lead to 
		\begin{equation}
		\rho_1(k_1) = \frac{1}{2\pi}\left(1 - \frac{u - v_1}{u - v_2}\right) + \rho_1(k_1).
		\end{equation}		
This equation has no solutions and hence clearly demonstrates that keeping only the reduced kernel $\tilde{K}$ is insufficient. We are thus forced to keep the $O(g^2)$ term in the kernel~\eqref{eqn:43} in order to establish the leading order behavior of the ground state densities in this regime.

Owing to this, the procedure for finding the zero temperature behavior is different from that employed in the previous section~\ref{linear}, where we first found the finite temperature solution and then took the limit $T\to0$. Here, we work directly at zero temperature. However, in order to find an analytically tractable solution, we make an ansatz that there is only a single Fermi momentum $k_i^f$ for the fermionic species $i$, which allows us to derive the ground state densities as a function of $k_i^f$. Given $\rho_i(k_i)$, we then calculate the particle densities $D_i$ and the total energy density $\mathcal{E}$ in terms of $k_i^f$. From these expressions, we then extract the chemical potential $h_i = \partial \mathcal{E}/\partial D_i$ and the inverse compressibility $\kappa_{ij}^{-1} = \partial h_i/\partial D_j$ to leading order in the coupling strength $g$. 
        
We note that while our results for physical quantities, such as the total energy, depend on the cut-off $\Lambda$, the compressibilities $\kappa_{ij}$ become cut-off independent in the limit $\Lambda\to\infty$. We also note that since we are unable to find an analytic solution when $v_1>0>v_2$, we will briefly discuss the failure of our ansatz in this region of parameter space at the end of this section.
        
We now study the case $v_1>u>v_2>0$ in detail and demonstrate the presence of a quasi-condensate phase and a quantum phase transition within this regime. Following the discussion above, we see that in order to find an analytic solution for the ground state densities in the weakly interacting limit, we are forced to keep the $O(g^2)$ term in the asymptotic expansion of the integral kernel $K(k_1,k_2)$. Thus, the TBA equations in the zero temperature limit are given by   
		\begin{align}
		\label{conddens}
        \rho_1(k_1) & = \frac{1}{2\pi} + \frac{v_1-u}{2 \pi} \int_{-\Lambda}^{k_2^f}dk_2\,\rho_2(k_2)K(k_1,k_2), \nonumber \\
        \rho_2(k_2) & = \frac{1}{2\pi} + \frac{u - v_2}{2 \pi} \int_{-\Lambda}^{k_1^f}dk_1\,\rho_1(k_1)K(k_1,k_2),
        \end{align}      
where $K$ is now given by Eq.~\eqref{eqn:43}. For momenta larger than the Fermi momentum, $k_i > k_i^f$, the densities vanish i.e., $\rho_i(k_i > k_i^f) =0$. 

The leading order term in the kernel $K(k_1,k_2)$ is still a $\delta$-function, which only acts (gives a non-zero contribution to the integrals above) if the momenta are in the range
        \begin{equation}
        \label{range1}
        \frac{-\Lambda(u - v_2) + \nu}{ v_1-u}<k_1<k_1^{(i)} ,
        \end{equation}
        for the first equation in Eq.~\eqref{conddens} and if
        \begin{equation}
        \label{range22}
        \frac{-\Lambda ( v_1-u) - \nu}{u - v_2}<k_2<k_2^{(i)} ,
        \end{equation}
for the second equation in Eq.~\eqref{conddens},
where we have defined 
		\begin{equation}
		\label{kint}
		\quad k_1^{(i)} = \frac{k_2^f (u - v_2) + \nu}{v_1-u},\quad k_2^{(i)} = \frac{k_1^f ( v_1-u) - \nu}{u - v_2}.
		\end{equation}		
For momenta outside this range (but within the Fermi sea), the $\delta$-function does not act and so the ground state densities are trivially given as 
		\begin{equation}
		\rho_1(k_1) = \frac{1}{2\pi} + O(g^2), \quad \rho_2(k_2) = \frac{1}{2\pi} + O(g^2),
		\end{equation}				
since the next term in $K(k_1,k_2) \propto g^2$. 		      

	\begin{figure}[t]
    \includegraphics[width=8cm]{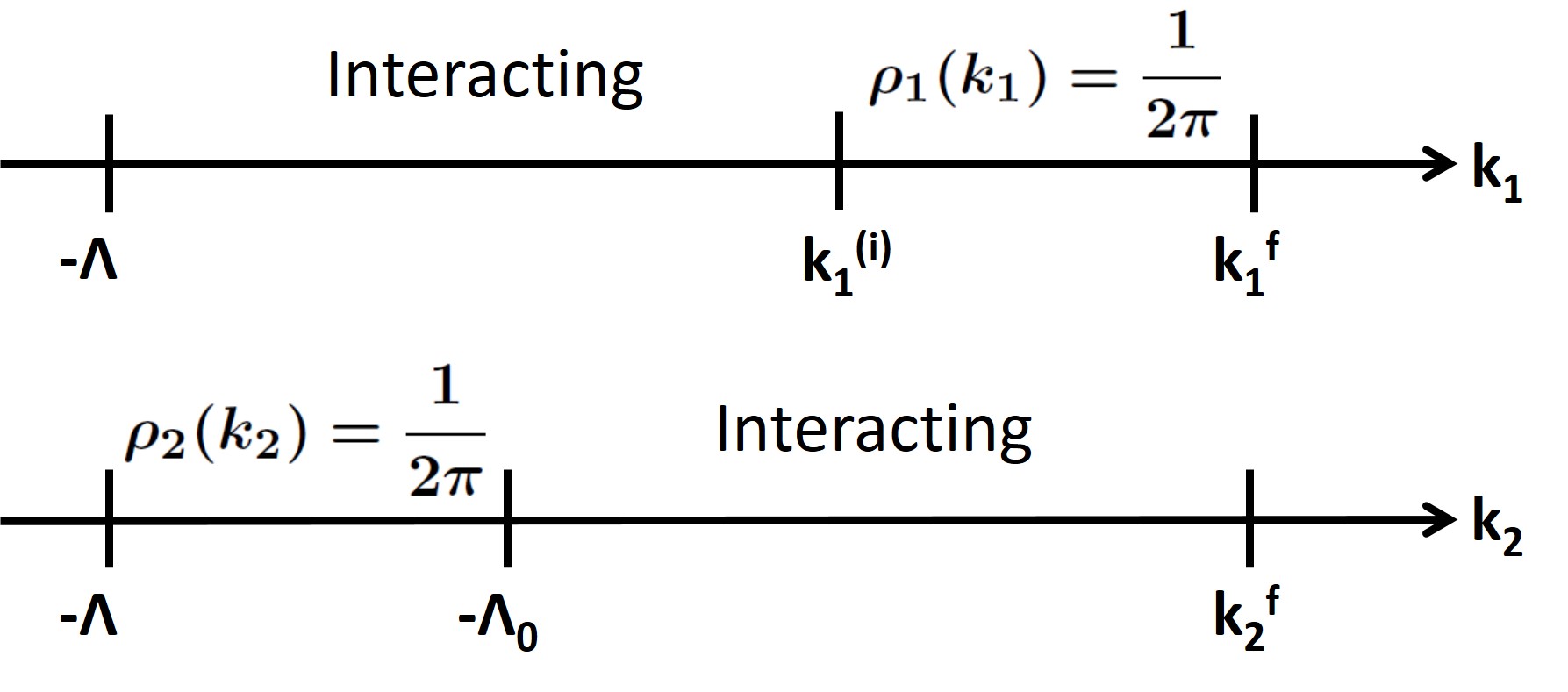}
    \centering
    \caption{For the case where $v_1 - u < u - v_2$ and $\alpha<$0, the interactions are restricted to lie between certain ranges of momenta, shown above. Outside these regions, the densities $\sim 1/(2\pi)$.}
    \label{fig:04} 
    \end{figure}
		      
In the following discussion, we assume without loss of generality that $v_1 - u < u - v_2$, since we find that the compressibilities are independent of this choice in the limit $\Lambda \to \infty$. In this case, Eqs.~\eqref{range1} and \eqref{range22}, which specify the range of momenta for which the $\delta$-function term in the kernel $K(k_1,k_2)$ contributes non-trivially to Eq.~\eqref{conddens}, can be replaced by 
		\begin{align}
		\label{range2}
		-\Lambda <&\,k_1 < k_1^{(i)}, \nonumber \\
		-\Lambda_0<&\,k_2 < k_2^{(i)},
		\end{align}			
where 
		\begin{equation}
		\Lambda_0 = \frac{\Lambda ( v_1-u) + \nu}{u - v_2}.
		\end{equation}
Thus, the particles of the first species will interact for all momenta up to the cut-off while the  particles of the second species will become non-interacting for large negative momenta $-\Lambda <k_2 < -\Lambda_0$.

It is now convenient to  introduce the parameter 
		\begin{equation}
		\label{alphacond}
		\alpha(k_1^f,k_2^f,\nu) = k_1^f (u - v_1) + k_2^f(u - v_2) + \nu, 
		\end{equation}
which also controls the range over which the particles interact. If we consider $\alpha<0$, then the  particles of the first species remain interacting only up to $k_1 = k_1^{(i)}$, while the particles of the second species interact all the way up to $k_2^f$. On the other hand, if $\alpha > 0$, then the first species of particles interact up to $k_1^f$ while the second species of particles only interact up to $k_2^{(i)}$. This is shown schematically for $\alpha < 0$ in Fig.~\ref{fig:04}.
    
We now elucidate the steps required for find the leading order solution for the ground state densities in the case where $\alpha<0$, with the the solutions for $\alpha = 0$ and $\alpha>0$ obtained by following the same procedure. As shown in Fig.~\ref{fig:04}, for momenta outside the interacting region i.e., where the $\delta$-function in the kernel does not act, the densities to leading order in $g$ are
        \begin{align}
        \label{rhononint}
        \rho_1(k_1) &= \frac{1}{2\pi}, &k_1^{(i)}<k_1<k_1^f ,\nonumber \\
        \rho_2(k_2) &= \frac{1}{2\pi}, &-\Lambda<k_2<-\Lambda_0.
        \end{align}
For momenta $k_i$ in the range set by Eq.~\eqref{range2}, however, the momentum densities satisfy
        \begin{widetext}
        \begin{align}
        	\rho_1(k_1) &= \frac{1}{2\pi} - \frac{u - v_1}{u - v_2} \rho_2\left(-\frac{k_1(u - v_1) + \nu}{u - v_2}\right) - g^2\frac{u - v_1}{2\pi (v_1 - v_2)} \int_{-\Lambda}^{k_2^f} dk_2\,\frac{\rho_2(k_2)}{(k_1(u-v_1) + k_2(u - v_2) + \nu)^2}, \nonumber \\
        	\rho_2(k_2) &= \frac{1}{2\pi} - \frac{u - v_2}{u - v_1} \rho_1\left(-\frac{k_2(u - v_2) + \nu}{u - v_1} \right) + g^2\frac{u - v_2}{2\pi (v_1 - v_2)} \int_{-\Lambda}^{k_1^f} dk_1\,\frac{\rho_1(k_1)}{(k_1(u-v_1) + k_2(u - v_2) + \nu)^2}.
        \end{align}
        \end{widetext}
In order to solve these equations, we break the integrals into the interacting and non-interacting regions, and replace $\rho_i(k_i)$ by its non-interacting value $1/(2\pi)$ in the latter.
 	\begin{figure}[t]
    \includegraphics[width=8cm]{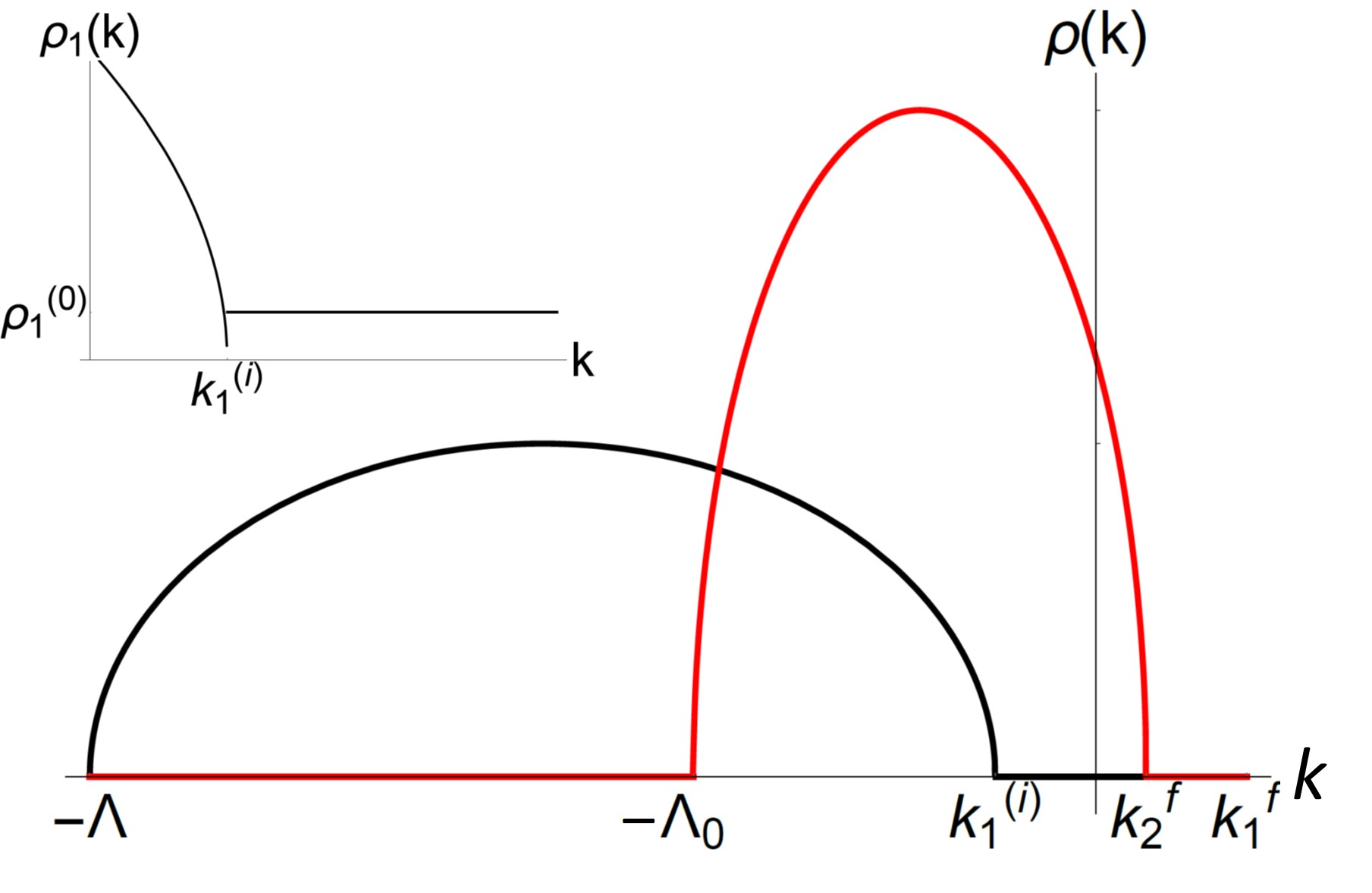}
    \centering
    \caption{\label{fig:05} The analytic solution for the ground state densities, $\rho_1(k_1)$ (in black) and $\rho_2(k_2)$ (in red) are plotted. Here $g = 0.1$, $v_1= 5$, $u = 4$, $v_1 = 2$, $\nu = - 20$, $\Lambda = 100$ , $k_1^f = 15$, and $k_2^f = 5$. These are shown for the case where $\alpha < 0$. The in-set shows $\rho_1(k_1)$ in the vicinity of $k_1^{(i)}$, where its behavior changes from interacting to non-interacting, with $\rho_1^{(0)} = 1/(2\pi)$ the non-interacting value. The discontinuity seen here is an artefact of the small $g$ approximation.}
    \end{figure}
    
As discussed before, we are keeping the $O(g^2)$ term in the integral kernel $K$. Keeping only the leading order $\sim 1/g^2$ terms, we find an integral equation for $\rho_1(k_1)$
        \begin{equation}
        \label{eqn:129}
        \int_{-\Lambda}^{k_1^{(i)}} dk'\,\frac{\rho_1(k')}{(k' - k)^2} = \frac{(v_1 - v_2)(u - v_1)}{2 g^2} \left(1 - \frac{u - v_1}{u - v_2} \right),
        \end{equation}
the solution to which is (see App.~\ref{app:E})
        \begin{equation}
        \label{rho1}
        \rho_1(k_1) = \frac{1}{2\pi g^2}\frac{(v_1 - u)(v_1 - v_2)^2}{u - v_2} \sqrt{(k_1 + \Lambda)(k_1^{(i)} - k_1)},
        \end{equation}
for $-\Lambda<k_1<k_1^{(i)}$. Similarly, we find that
        \begin{equation}
        \label{rho2}
        \rho_2(k_2) = \frac{1}{2\pi g^2}\frac{(u - v_2)(v_1 - v_2)^2}{|u - v_1|} \sqrt{(k_2^f - k_2)(k_2 + \Lambda_0)},
        \end{equation}
for $-\Lambda_0<k_2<k_2^f$. Thus, we find qualitatively different behavior in this parameter regime, since unlike the densities found in Sec.~\ref{linear}, which were discontinuous momentum-independent constants, the ground state densities found here are momentum dependent and diverge in the limit $g\to 0$. Note the striking resemblance of this solution to the behavior of the weakly interacting Bose gas (Lieb-Liniger model \cite{LiebLiniger}). Hence the conclusion that  the weakly interacting regime of the Q3WI with velocities such that $v_1>u>v_2>0$ describes what we term a quasi-condensate phase. 

These ground state densities $\rho_i(k_i)$ are illustrated in Fig.~\ref{fig:05}. In order to verify that our solution indeed satisfies the TBA equations, we numerically substitute our solution into the integral equations~\eqref{conddens} but now with the complete integral kernel Eq.~\eqref{eqn:18} and match the left and right hand sides of those equations. We find excellent agreement between the two for $g\ll 1$, hence validating our analytic solution. 
   
It is important to note that our solution, given in Eqs.~\rf{rho1} and \rf{rho2}, breaks down at the end points, as it predicts that the densities vanish there as an artefact of the small $g$ approximation, whereas in reality the densities smoothly connect with the $1/(2\pi)$ densities of the non-interacting regions given by Eq.~\rf{rhononint}.

Due to this, unlike in Sec.~\ref{linear} where we could find an expression for the excitation spectra $\epsilon_i$ and demand that they vanish at the Fermi-momenta in order to find the chemical potentials $h_i$, here our expressions for $\epsilon_i$---obtained by solving integral equations similar to those for the densities---will always vanish at the end points for any $h_i$. Thus, in a way which parallels the case of the weakly interacting Lieb-Liniger gas, we cannot obtain $h_i$ by finding the excitation spectra but must use the method described at the beginning of this section, wherein we must find the particle densities $D_i$ and total energy density $\mathcal{E}$ as functions of the Fermi momenta $k_i^f$ in order to derive the chemical potentials and inverse compressibility. 


To implement this method, we compute  the particle densities 
\beq
\label{dens}
D_1 = \int_{-\Lambda}^{k_1^f} dk_1\, \rho_1(k_1),\quad D_2 = \int_{-\Lambda}^{k_2^f} dk_2\, \rho_2(k_2),
\eeq 
where the densities $\rho_i$ are given by Eqs.~\eqref{rhononint},~\eqref{rho1}, and ~\eqref{rho2}. These integrals can be evaluated analytically in order to find the particle densities as functions of the Fermi momenta, i.e. $D_i = D_i(k_1^f,k_2^f)$. Similarly, we can express the total energy
\beq
\label{energy}
\mathcal{E} = v_1 \int_{-\Lambda}^{k_1^f} dk_1 \,k_1 \rho_1(k_1) + v_2 \int_{-\Lambda}^{k_2^f} dk_2 \,k_2 \rho_2(k_2)
\eeq
as a function of the Fermi momenta, $\mathcal{E} = \mathcal{E}(k_1^f,k_2^f)$. The chemical potentials are then found from the derivatives
\begin{equation}
\label{eq:chempot}
h_i=\frac{\partial \mathcal{E}}{\partial D_i} = \sum_{n = 1,2} \frac{\partial k_n^f}{\partial D_i} \frac{\partial \mathcal{E}}{\partial k_n^f}.
\end{equation}
Differentiating once more allows us to find the inverse compressibilities. 
Importantly, we take the limit $\Lambda \to \infty$ in order to find the leading order (in $g$) compressibilities, which are cut-off independent,
\begin{align}
\label{eq:ccco}
\left[\kappa \right]^{-1}_{ij} &= \lim_{\Lambda \to \infty} \frac{\partial^2 \mathcal{E}}{\partial D_i \partial D_j} \nonumber \\
&= \lim_{\Lambda \to \infty} \sum_{m,n=1,2} \frac{\partial k_m^f}{\partial D_i} \frac{\partial}{\partial k_m^f} \left( \frac{\partial k_n^f}{\partial D_j} \frac{\partial \mathcal{E}}{\partial k_n^f} \right)
\end{align}

Before we proceed with the calculation, we must justify our ansatz that the excitation spectra cross zero-energy only once. In order to do this, we must check that the excitation energies remain negative for all $k_i \leq k_i^f$ since by definition, $\epsilon_i(k_i) < 0$ when $\rho_i(k_i) > 0$. This will demonstrate that our solutions are consistent with our initial assumption that $\epsilon_i(k_i) = 0$ only at a single point. 

To find the excitation spectra at zero temperature in the weakly interacting limit, we must solve the coupled integral equations
		\begin{align}
		\label{eq:coupledspec}
		\epsilon_1(k_1) & =  k_1 v_1 - h_1 + \frac{u - v_2}{2\pi} \int_{-\Lambda}^{k_2^f} dk_2 \,K(k_1, k_2) \, \epsilon_2( k_2), \nonumber \\
		\epsilon_2(k_2) & =  k_2 v_2 - h_2 - \frac{u - v_1}{2\pi} \int_{-\Lambda}^{k_1^f} d k_1\,K(k_1,k_2) \,  \epsilon_1(k_1),
		\end{align}
with the kernel given by Eq.~\eqref{eqn:43}. 

The solution proceeds in a manner similar to that for the densities, as in the range of momenta where the $\delta$-function does not act, we find that (for $\alpha < 0$) 
		\begin{align}
		\label{epsnonint}
		\epsilon_1(k_1) & = k_1 v_1 - h_1, \quad k_1^{(i)} <k_1 < k_1^f, \nonumber \\
		\epsilon_2(k_2) & = k_2 v_2 - h_2, \quad - \Lambda <k_2 < -\Lambda_0.
		\end{align}
On the other hand, for the excitation spectra in the interacting region, we must solve integral equations following from Eqs.~\rf{eq:coupledspec}, similarly to how it was done for the 
densities. We find that the leading order $\sim 1/g^2$ solutions for these are
        \begin{align}
        \epsilon_{1}&(k_1) = \frac{(v_1 - u)( v_1 - v_2)}{g^2} \sqrt{(k_1^{(i)} - k_1)(k_1 + \Lambda)} \nonumber \\ 
        &\times \left( u \frac{v_1 - v_2}{u - v_2}\left(\frac{k_1^{(i)} - \Lambda + 2 k_1}{4} \right) -h_1 - h_2 - \frac{\nu v_2}{u - v_2}\right),
        \end{align}
for $-\Lambda<k_1<k_1^{(i)}$, and        
        \begin{align}
        \epsilon_{2}&(k_2) = \frac{(u - v_2)( v_1 - v_2)}{g^2} \sqrt{(k_2^f - k_2)(k_2 + \Lambda_0)} \nonumber \\ 
        & \times \left( u \frac{v_1 - v_2}{v_1 - u}\left(\frac{k_2^f - \Lambda_0 + 2 k_2}{4} \right) -h_1 - h_2 - \frac{\nu v_1}{v_1 - u}\right),
        \end{align}
for $-\Lambda_0<k_2<k_2^f$. 
While we found these solutions for arbitrary $h_i$, in practice those must be fixed according to Eq.~\rf{eq:chempot}. 

\begin{figure}[t]
    \includegraphics[width=8cm]{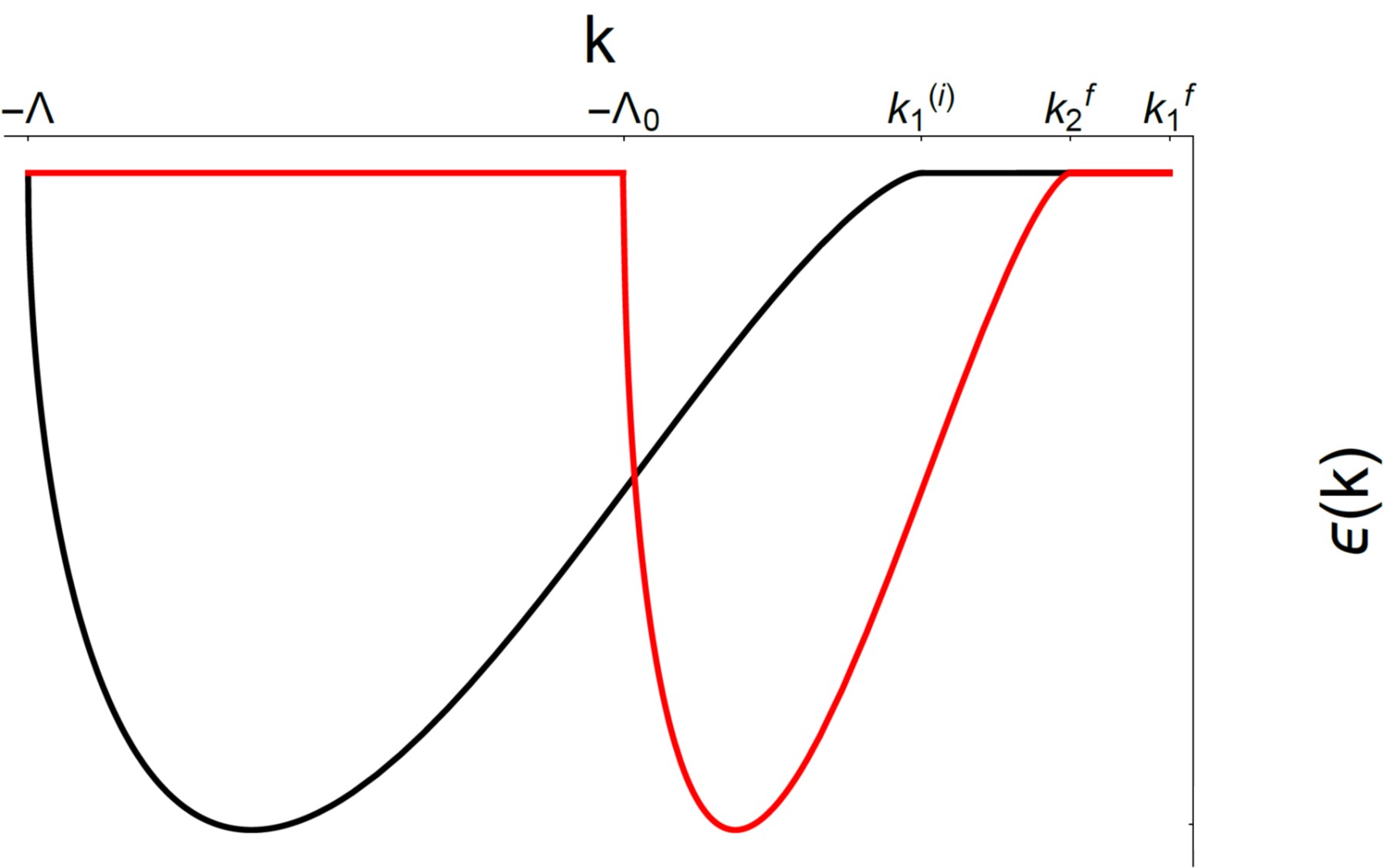}
    \centering
    \caption{\label{fig:06} The analytic solution for the excitation energies, $\epsilon_{1}(k_1)$ (in black) and $\epsilon_{2}(k_2)$ (in red) are plotted. Here $g = 0.1$, $v_1 = 5$, $u = 4$, $v_1 = 2$, $\nu  = -20$, $\Lambda  = 100$, $k_1^f = 15$, and $k_2^f = 5$. This is for the case where $\alpha < 0$. For $k_i > k_i^f$, the excitation spectra $\epsilon_i(k_i) >0$, hence confirming our ansatz that the spectra only cross zero energy once.}
    \end{figure}

For the case $\alpha<0$, with the chemical potentials appropriately fixed,  the excitation spectra are plotted in Fig.~\ref{fig:06}.

Although we have only stated solutions for $k_i < k_i^f$ here, the integral equations~\eqref{eq:coupledspec} can also be solved for momenta larger than the Fermi momenta i.e., $k_i > k_i^f$. We have verified that indeed $\epsilon_i(k_i>k_i^f) >0$, hence vindicating our ansatz that the spectra only cross zero energy at a single point. 	 
        


Now that we have constructed the solution to our problem, we would like to proceed with calculating the compressibilities to verify that they change when $\alpha$ changes its sign, indicating a phase transition
when $\alpha=0$. 
At first glance, given that we have derived analytic expressions for the ground state densities, it appears straightforward to derive analytic expressions for the particle densities $D_i$~\eqref{dens} and the total energy $\mathcal{E}$~\eqref{energy} and to then evaluate the derivatives in Eq.~\eqref{eq:ccco} to determine compressibilities. 

However, as discussed earlier, our expressions for the ground state densities $\rho_i$ in the interacting regions, given by Eqs.~\eqref{rho1},~\eqref{rho2}, break down in the vicinity of the end points i.e., momenta where the behavior changes from interacting to non-interacting. Since the true behaviour of the densities is expected to be one where they connect smoothly with the non-interacting value $1/(2\pi)$, we must account for the possibility that perturbative in $g$ corrections to our solutions lead to non-trivial contributions to the compressibilities. Hence, care must be taken while evaluating the derivatives that appear in Eq.~\eqref{eq:ccco}, a detailed discussion of which is presented in App.~\ref{app:F}.

The analysis presented in the Appendix shows that any corrections to our expressions for the ground state densities will in fact not contribute to the leading order (in $g$) ground state inverse compressibilities, which we find are  
\begin{align}
\label{comps}
\kappa^{-1} &= 2 \pi v_1
\left( 
\begin{array}{cc}
1 & -1 \\
-1 & 1
\end{array}
\right),\quad \alpha<0, \nonumber \\
\kappa^{-1} &= 2 \pi v_2
\left( 
\begin{array}{cc}
1 & -1 \\
-1 & 1
\end{array}
\right),\quad \alpha>0.
\end{align}
We first note that the cut-off, $\Lambda$, drops out entirely in these expressions since we are working in the limit $\Lambda \to \infty$. We also observe that since we are working in the regime where $v_1>u>v_2>0$, the inverse compressibilities are well-defined and positive. 

The situation is somewhat more involved for the compressibilities themselves. We observe that the inverse compressibility matrices in either regime are non-invertible. This is another artifact of the small $g$ approximation since it is possible to show (see App.~\ref{app:F}) that these matrices are necessarily invertible once corrections to the leading order solutions are accounted for. Thus, in principle it is possible to find the compressibility matrix $\kappa$ by extending our analytic solutions for the densities $\rho_i$ beyond leading order.

The inverse compressibility matrices~\eqref{comps} suffice, however, for the purposes of establishing a phase transition within the quasi-condensate regime since we see that $\kappa^{-1}$ is discontinuous as a function $\alpha$. In particular, $\kappa^{-1}$ equals the non-interacting compressibility of the first fermionic species if $\alpha<0$, and of the second fermionic species if $\alpha>0$. This can be understood as follows: for $\alpha<0$, the majority $v_1$ particles are non-interacting in the vicinity of their Fermi momentum $k_1^f$ since $\rho_1(k_1) = 1/(2\pi)$ here. Since the compressibility depends primarily on the behavior close to the Fermi point, we expect it to take on its non-interacting value for the $v_1$ particles.

Thus, if $\alpha<0$, only particles of the first species contribute to the compressibility of both species. The situation is reversed when $\alpha>0$ since the compressibility is now determined by the second species of fermions, which are non-interacting in the vicinity of $k_2^f$. 

Just as was done in the prior study of Case A in Sec.~\ref{linear}, we can re-express the parameter $\alpha$ in terms of the densities $D_1$ and $D_2$, to get a clear criterion of the transition as a function
of densities. It is straightforward to check by integrating (\ref{rho1}) and (\ref{rho2})  that the contribution of the interacting regions to $D_1$ is exactly equal to their contribution to $D_2$, and hence
\begin{equation} D_2 - D_1  = \frac 1 {2 \pi} \left(  \left(\Lambda- \Lambda_0 \right) -  \left( k_1^f - k_1^{(i)} \right) \right).
\end{equation}
In turn, we rewrite this with the help of Eq.~(\ref{kint}) as
\begin{equation} D_2 - D_1 = \frac{\alpha}{2 \pi  (v_1-u)} + \frac{\Lambda-\Lambda_0}{2\pi}.
\end{equation}
Therefore we see that since the transition happens when $\alpha$ changes sign, this is equivalent to saying that the transition happens when
\begin{equation} \label{eq:betab} \beta = D_2 - D_1 - (\Lambda-\Lambda_0)/(2\pi) \end{equation} changes sign. Since the last term in Eq.~(\ref{eq:betab}) is a density independent constant, this gives a clear criterion for transition as a function of the densities. 
      
When $v_1>0>v_2$ (with $v_1>u>v_2$) we are unable to find a consistent analytic solution for the ground state densities and excitation spectra. In particular, our ansatz that the excitation spectra cross zero energy at only one point breaks down since making that ansatz leads to negative compressibilities, indicating the instability of our solution. Additionally, if we construct the excitation spectrum for this case following our steps in the previous sections, we find that it is positive where the densities are positive: another contradiction, since by definition, $\epsilon(k) < 0$ where $\rho(k) > 0$. Thus, our ansatz that there is only one Fermi-point breaks down here. This is similar to the case in Sec.~\ref{sec:A2b} where the excitation spectrum changes sign twice. While we did find some numerical evidence for this being the case, our solution did not converge and we will hence leave this case for future work.  


\section{Conclusions}
\label{cncls}

In this paper, we have studied a one-dimensional model of chiral fermions interacting through an $s$-wave Feshbach resonance. In particular, we have demonstrated that this model is host to a plethora of possible behaviors in the weakly interacting regime, from one where the fermions remain interacting deep within their Fermi seas to one where they become non-chiral. We also find that fermions interacting via a Feshbach resonance, in the limit of weak interactions, can exhibit behavior that closely resembles that of the weakly interacting Bose gas, a case that we have dubbed the ``quasi-condensate" phase of our model.

While we have demonstrated the existence of a quantum phase transition through the discontinuity in the zero temperature compressibility, the precise nature of this transition remains unclear and may benefit from a bosonization study. It would also be interesting to further investigate the thermodynamic properties of the phases where the particles interact deep into the Fermi sea---is this unusual behavior reflected in any quantities besides the compressibility?

The full phase diagram of our system as a function of the interaction strength $g$ and of the particle densities remains the subject of future work, since we can only probe the weak $g$ and strong $g$ limits analytically. Studying generic values of $g$ would probably require a numerical study of the Bethe Ansatz equations derived here.

We were unable to find an analytic solution in one case, that when $v_1>0>v_2$ and the molecular velocity $u$ lies in between the atomic velocities $v_i$. In order to complete our characterization of this model, we must thus develop a numerical solution for the Q3WI model in this particular regime. However, from our analysis of the other cases, we expect that this particular regime should also host a quasi-condensate phase and exhibit a quantum phase transition as a function of particle densities. \\

\paragraph*{Acknowledgements:} We are grateful to Will Jay, Alex Kamenev, Sergej Moroz, and Yang-Zhi Chou for useful discussions. We acknowledge financial support from NSF grants DMR-1205303 and PHY-1211914.


\newpage
\appendix

\section{$T$-Matrix calculation of scattering amplitudes}
	\label{app:A}

Here, we calculate the scattering amplitude of an $a_1$ particle and an $a_2$ particle governed by the Q3WI Hamiltonian Eq.~\eqref{q3wi}. This is done by calculating the amplitude for the diagram 

\begin{fmffile}{diagram}
		\begin{eqnarray*}	
		\parbox{25mm}{\begin{fmfgraph}(3cm,1.5cm)
			\fmfleftn{i}{2}
			\fmfrightn{o}{2}
			\fmf{fermion}{i1,v1}
			\fmf{fermion}{i2,v1}
			\fmf{photon}{v1,v2}
			\fmfblob{.15w}{v2}
			\fmf{photon}{v2,v3}
			\fmf{fermion}{v3,o1}
			\fmf{fermion}{v3,o2}
			\fmflabel{$\frac{p}{2} + k$}{i1}
			\fmflabel{$\frac{p}{2} - k$}{i2}
			\fmflabel{$\frac{p}{2} + k'$}{o1}
			\fmflabel{$\frac{p}{2} - k'$}{o2}
		\end{fmfgraph}} & = &  \parbox{25mm}{\begin{fmfgraph}(3cm,1.5cm)
\fmfleftn{i}{2} \fmfrightn{o}{2}
			\fmf{vanilla}{i1,v1}
			\fmf{vanilla}{i2,v1}
			\fmf{photon}{v1,v2}
			\fmf{vanilla}{v2,o1}
			\fmf{vanilla}{v2,o2}
\end{fmfgraph}} + \parbox{25mm}{\begin{fmfgraph}(3cm,1.5cm)
\fmfleftn{i}{2} \fmfrightn{o}{2}
			\fmf{vanilla}{i1,v1}
			\fmf{vanilla}{i2,v1}
			\fmf{photon}{v1,v2}
			\fmf{vanilla,left,tension=.5}{v2,v3}
			\fmf{vanilla,right,tension=.5}{v2,v3}
			\fmf{photon}{v3,v4}
			\fmf{vanilla}{v4,o1}
			\fmf{vanilla}{v4,o2}
\end{fmfgraph}} + \\ & & \parbox{25mm}{\begin{fmfgraph}(4cm,1.5cm)
\fmfleftn{i}{2} \fmfrightn{o}{2}
			\fmf{vanilla}{i1,v1}
			\fmf{vanilla}{i2,v1}
			\fmf{photon}{v1,v2}
			\fmf{vanilla,left,tension=.5}{v2,v3}
			\fmf{vanilla,right,tension=.5}{v2,v3}
			\fmf{photon}{v3,v4}
			\fmf{vanilla,left,tension=.5}{v4,v5}
			\fmf{vanilla,right,tension=.5}{v4,v5}
			\fmf{photon}{v5,v6}
			\fmf{vanilla}{v6,o1}
			\fmf{vanilla}{v6,o2}
\end{fmfgraph}} \hspace{1.5cm} + ...
\end{eqnarray*} 
\end{fmffile}

Since these diagrams form a geometric series, they can be re-summed exactly to give the $T$-matrix amplitude
	\begin{equation}
	T(k,E) = \frac{g^2}{D^{-1}(k,E) - g^2 \Pi(k,E)} ,
	\end{equation}
where
\begin{fmffile}{diagram2}
	\begin{eqnarray*}	
		\parbox{20mm}{\begin{fmfgraph}(1.5cm,1.5cm) \fmfleft{i}\fmfright{o}\fmf{photon}{i,o}\end{fmfgraph}} & = & D(k,E),\\
		\parbox{20mm}{\begin{fmfgraph}(1.5cm,1.5cm) \fmfleft{i}\fmfright{o}\fmf{fermion,left,tension=0.5}{i,o}\fmf{fermion,right,tension=0.5}{i,o}\end{fmfgraph}} & = & \Pi(k,E).
\end{eqnarray*}	
\end{fmffile}
Here, $D$ is the retarded Green's function for the $b$-particles
		\begin{equation}
		D(k,E) = \frac{1}{E - u k - \nu + i 0} ,
		\end{equation}
and $\Pi$ is the polarization operator
        \begin{equation}
        \Pi = \int \frac{d\omega}{2\pi i}\frac{dk}{2\pi} G_{0,1}\left( \frac{p}{2} + k, E + \omega \right) G_{0,2} \left( \frac{p}{2} - k, - \omega \right) ,
        \end{equation}
where $G_{0,i}$ is the retarded Green's function for the $a_i$ particles
		\begin{equation}
		G_{0,i}(k,E) = \frac{1}{E - v_i k + i 0} ,
		\end{equation}
and $p = k_1 + k_2$ is the centre of mass momentum. We find that
        \begin{align}
        \Pi &= \int_{-\infty}^{\infty} \frac{dk}{2\pi} \frac{1}{E - \frac{p}{2}(v_1 + v_2) - k(v_1 - v_2) - i \epsilon } , \nonumber \\ &= -\frac{i}{2 (v_1 - v_2)} .
        \end{align}
Since the energy is given by $E = k_1 v_2 + k_2 v_2$, we find
        \begin{equation}
		T = \frac{g^2}{\frac{i g^2}{2(v_1 - v_2)} + k_1 v_1 + k_2 v_2 -  u (k_1 + k_2) - \nu} .
		\end{equation}
To calculate the $S$-matrix from the $T$-matrix, we note that the scattering amplitude $f$ equals $\frac{S - 1}{2}$ and for a linear dispersion satisfies~\cite{Gurarie2006}
		\begin{align}
		\label{eqn:A7}
		f(k_1,k_2) &= -i\,\frac{T(k_1,k_2)}{2 (v_1 - v_2)} ,\nonumber \\
		&= \frac{-i g^2}{i g^2 + 2 \left(v_1 - v_2\right) \left(k_1 (v_1 - u) + k_2 (v_2 - u )- \nu \right)}.
		\end{align}
		

	\section{Linearisation of the continuum two-channel model}
	\label{app:B}
	
Here, we relate the scattering amplitude for the two-channel Hamiltonian Eq.~\eqref{eqn:01}, to that of the Q3WI Hamiltonian by linearising the two-channel model in the vicinity of certain Fermi points. We first calculate the $T$-matrix for Eq.~\eqref{eqn:01}, following the method outlined in App.~\ref{app:A},
		\begin{equation}
		T = \frac{g^2}{\frac{i g^2 m_a}{k_1 -k_2}+\frac{k_1^2}{2 m_a}+\frac{k_2^2}{2 m_a}-\frac{\left(k_1+k_2\right){}^2}{2 m_b}-\epsilon _0} .
		\end{equation}		
Now, for quadratic dispersions, the scattering amplitude $f$ is related to the $T$-matrix as
		\begin{equation}
		f(k) = - i \frac{\mu}{k} T ,
		\end{equation}
where $\mu = \frac{m}{2}$ is the reduced mass, and $k = \frac{k_1 - k_2}{2}$ is the reduced momentum. This leads to 
		\begin{equation}
		f(k_1,k_2) = \frac{-i g^2}{\left(\frac{k_1}{m_a}-\frac{k_2}{m_a}\right) \left(\frac{k_1^2}{2 m_a}+\frac{k_2^2}{2 m_a}-\frac{\left(k_1+k_2\right){}^2}{2 m_b}-\epsilon _0\right)+i g^2} .
		\end{equation}
For ease of notation, we define						
		\begin{equation}
		t_i \equiv \frac{1}{2 m_i} ,\quad i = a \,,b 
		\end{equation}				
such that the scattering amplitude equals
		\begin{equation}
		f = \frac{-i g^2}{2 \left(k_1 t_a-k_2 t_a\right) \left(k_1^2 t_a+k_2^2 t_a-\left(k_1+k_2\right){}^2 t_b-\epsilon _0\right)+i g^2}	 .
		\end{equation}
We now linearise the spectrum about two Fermi-momenta,
		\begin{equation}
		k_1 \approx k_{f1} + x_1 ,\quad k_2 \approx k_{f2} + x_2 .
		\end{equation}									
To linear order in $x_i/k_{fi}$, we thus find
		\begin{equation}
		 f \approx \frac{-i g^2}{i g^2 + 2 \left(v_1 - v_2\right) \left(x_1 v_1 + x_2 v_2 - u (x_1 + x_2) - \nu \right)} ,
		\end{equation}
where we have defined the parameters
		\begin{align}
		v_1 & = \frac{\sqrt{t_a}}{d \sqrt{2}} \left( (k_{f1} - k_{f2})^2 t_a - (k_{f1} + k_{f2})^2 t_b + 2 k_{f1}^2 t_a - \epsilon_0 \right) ,\nonumber \\
		v_2 & = \frac{-\sqrt{t_a}}{d \sqrt{2}} \left( (k_{f1} - k_{f2})^2 t_a - (k_{f1} + k_{f2})^2 t_b + 2 k_{f2}^2 t_a - \epsilon_0 \right) ,\nonumber \\
		u & = \frac{2 t_b}{d} \sqrt{\frac{t_a}{2}} \left(k_{f1}^2 - k_{f2}^2 \right) ,\nonumber \\
		\nu & =  \frac{(k_{f1} - k_{f2})\sqrt{t_a}}{d\sqrt{2}} \left( (k_{f1}^2 + k_{f2}^2)t_a - (k_{f1} + k_{f2})^2 t_b - \epsilon_0\right) ,
		\end{align}
and
		\begin{equation}
		d = \pm \sqrt{2 t_a \left(k_{f1}^2 - k_{f1} k_{f2} + k_{f2}^2\right) - t_b (k_{f1} + k_{f2})^2 -\epsilon_0} .
		\end{equation}				
Taking the positive root for $d$ leads to $v_1 > v_2$, while the negative root results in $v_2 > v_1$.

In the continuum, we require that $m_b = 2 m_a$, which pins the molecular velocity to that of the atoms, $u = (v_1  + v_2)/2$. However, on a lattice the effective mass of the molecule can be made independent of that the atoms, in which case $u$ can be varied (see App.~\ref{app:C}). Furthermore, in the case that $k_{f1} = k_{f2}$, we find that $v_2 = - v_1$ and $u = 0$. This corresponds to the case $v_1>0>v_2$ with $u$ lying in between, where unfortunately we are unable to find an analytic solution for the ground state densities (this is discussed in Sec.~\ref{condensate}).

We note that to reach some of the regimes, for instance $u>v_1>0>v_2$, we require a negative molecular mass, $m_b < 0$, which while not possible in the continuum, is allowed on a lattice. While the square root in the parameter $d$ forces the original parameters $m_a, m_b, k_{f1}, k_{f2}, \nu$ to be chosen such that the velocities remain real, we will see that this linearisation procedure, when applied to the lattice two-channel model, allows us to access all the regimes we study in the main text.

  
	\section{Linearisation of the lattice two-channel model}
	\label{app:C}
	
		On a 1D lattice, the two-channel model is defined as 
		\begin{align}
		H = & - t_a \sum_{\braket{r , r'}} c_{r,\sigma}^{\dagger} c_{r',\sigma}  - t_b \sum_{\braket{r , r'}} b_{r}^{\dagger} b_{r'} + \epsilon_0 \sum_r b_r^\dagger b_r \nonumber \\& + g \sum_r (b_r^\dagger c_{r,\uparrow} c_{r,\downarrow} + h.c.) ,
		\end{align}			
	where $c_{r,\sigma},c_{r,\sigma}^\dagger $ are annihilation and creation operators for fermionic atoms with spin $\sigma$ at the lattice site $r$, and $b_r,b_r^\dagger$ are annihilation and creation operators for closed channel bosonic molecules. $\epsilon_0$ is the bare detuning, $g$ is the coupling, $t_a (t_b)$ are hopping matrix elements for atoms (molecules), and $\braket{r,r'}$ denotes nearest neighbour interactions. In the absence of interactions, the atoms and molecules are free particles with lattice dispersions,
		\begin{equation}
		\epsilon_k = - 2 t \text{ cos(k)}  ,
		\end{equation}		 
		\begin{equation}
		\epsilon_k^b = - 2 t_b \text{ cos(k)} + \epsilon_0 ,
		\end{equation}
with the lattice constant, $a$, set to 1.

This model in three dimensions was studied in \cite{Shen2012}, using a mean-field analysis, where it was shown that when the band is more than half-filled and as the detuning is decreased (equivalently, attractive interactions are increased), the gas exhibits a non-monotonic crossover, from a paired BCS superfluid to a BEC of molecular holes, to the BCS, and to the BEC of molecules.

The scattering amplitude for this model can be calculated as described in App.~\ref{app:A} (where the momentum integration in the polarization operator $\Pi$ now goes over the first Brillouin-zone, $k\in[-\pi,\pi]$),
		\begin{widetext}
		\begin{equation}
		f(k_1,k_2) = \frac{- i g^2}{i g^2 + 2 t_a \Big(\sin(k_1) - \sin(k_2) \Big)\Big( 2 t_b \cos(k_1 + k_2) - 2 t_a \left(\cos(k_1) + \cos(k_2) \right)- \epsilon_0 \Big)} .
		\end{equation}
		\end{widetext}
Similar to the continuum case, in the vicinity of two distinct Fermi momenta, $k_1 \approx k_{f1} + x_1$ and $k_2 \approx k_{f2} + x_2$, this scattering amplitude reduces to that for the Q3WI (Eq.~\eqref{eqn:A7}), where the velocities are now defined as
		\begin{widetext}
		\begin{align}
		v_1 & = -\frac{\sqrt{t_a}}{d} \Bigg(2 t_a \Big( \cos(2 k_{f1}) + \cos(k_{f1} - k_{f2}) \Big) - \cos(k_{f1})\Big( 2 t_b \cos(k_{f1} + k_{f2}) - \epsilon_0 \Big) \Bigg) ,\nonumber \\
		v_2 & = \frac{\sqrt{t_a}}{d} \Bigg(2 t_a \Big( \cos(2 k_{f2}) + \cos(k_{f1} - k_{f2}) \Big) - \cos(k_{f1})\Big( 2 t_b \cos(k_{f1} + k_{f2}) - \epsilon_0 \Big) \Bigg) ,\nonumber \\
		u & = 2 t_b \frac{\sqrt{t_a}}{d} \sin(k_{f1} + k_{f2}) \Big(\sin(k_{f1}) - \sin(k_{f2}) \Big), \nonumber \\
		\nu & = -\frac{\sqrt{t_a}}{d} \Bigg(2 t_a \Big(\cos(k_{f1}) + \cos(k_{f2}) \Big) - 2 t_b \cos(k_{f1} + k_{f2}) + \epsilon_0\Bigg)\Big(\sin(k_{f1}) - \sin(k_{f2}) \Big) ,\nonumber \\
		d & = \pm \sqrt{\Big( \cos(k_{f1}) + \cos(k_{f2}) \Big)\Big( 2 t_b \cos(k_{f1} + k_{f2}) - \epsilon_0 \Big) - 2 t_a \Big( \cos(2 k_{f1}) + 2 \cos(k_{f1} - k_{f2}) + \cos(2 k_{f2})  \Big)} .
		\end{align}
		\end{widetext}
Taking the positive root for $d$ leads to $v_1 > v_2$, while the negative root leads to the regime where $v_2 > v_1$. Thus, by varying the Fermi-points, $k_{f1}$ and $k_{f2}$, the relative strengths of the atomic and molecular hoppings $t_a\,,t_b$, and the detuning $\epsilon_0$, we can explore all parameter regimes of the Q3WI model. Moreover, on the lattice we have more freedom as now $t_a$ and $t_b$ are independent, unlike in the continuum case. 
	 
	 
	\section{Details for weakly interacting regime}
	\label{app:D}
	
In Sec.~\ref{linear}, we have discussed the general procedure for deriving the zero temperature ground state densities $\rho_i(k_i)$ and excitation spectra $\epsilon_i(k_i)$ in the weakly interacting regime for the cases when $u>v_1>v_2$. Here, we present in detail the ground state densities, excitation spectra, and compressibilities for these cases. We reiterate that a host of other cases may be obtained from the ones discussed here; in particular, cases where $v_1>v_2>u$ can be obtained from the cases studied here by a combination of parity and fermion type exchange. 

		\subsection{Case A: $u > v_1 > v_2$}
		\label{caseA}

For this case, the excitation spectra and ground state densities are found by taking the limit $T\to0$ in Eq.~\eqref{eqn:47} and Eq.~\eqref{density} respectively. The special momenta $k_i^{(j)}$ ($j=a,b,c$) at which the derivatives $\epsilon'_i(k_i)$ may exhibit discontinuities are defined in Eq.~\eqref{parameters}. We note that case A1, which corresponds to $u>v_1>v_2>0$, was analyzed in the main text and we will hence not repeat its discussion here. 

        \subsubsection{Case A2i: $u>v_1>0>v_2$ \& $u < \frac{2 v_1 v_2}{v_1 + v_2}$}  
        \label{sec:A2a}
For this case, we find that the parameter $\beta$, which satisfies $\text{sgn}(\beta) = \text{sgn}(\alpha)$, is  
        \begin{align}
        \beta(D_1,D_2) = &4 \pi  \left(\frac{D_2}{2 u-v_1-v_2}-\frac{D_1}{v_1-v_2}\right) \nonumber \\
        & -\frac{\nu }{\left(u-v_1\right) \left(u-v_2\right)}.
        \end{align}
The excitation spectra for this case are shown in the main text, Fig.~\ref{fig:02}. Depending on $\text{sgn}(\beta)$, the ground state densities and compressibilites are as follows: 
        
For $\beta<0$, the ground state densities are
        \begin{align}
        &\rho_1(k_1) = 
        \begin{dcases}
        \frac{1}{4\pi} \left(1 - \frac{u - v_1}{u - v_2} \right), &k_1<k_1^{(a)}, \\
        \frac{1}{2\pi}, &k_1^{(a)}<k_1<k_1^f,\\
        0, &k_1>k_1^f,
        \end{dcases}\nonumber \\
        &\rho_2(k_2) = 
        \begin{dcases}
        0, &k_2<k_2^{(b)},\\
        \frac{1}{4\pi} \left(1 + \frac{u - v_2}{u - v_1} \right), &k_2>k_2^{(b)},
        \end{dcases}
        \end{align}
where the Fermi momenta are
		\beq
		k_1^f = \frac{h_1}{v_1},\quad k_2^f = k_2^{(b)}.
		\eeq
These lead to the compressibility
        \begin{equation}
        \kappa^{-1} = 2 \pi \left(
        \begin{array}{cc}
        v_1 & v_1\\
        v_1 & \frac{3 v_1 v_2 - 2 u v_2 - v_1^2}{2u - v_1 - v_2}
        \end{array} \right). 
        \end{equation}
We note that the compressibilities are positive as required on physical grounds. In particular, $3 v_1 v_2 - 2 u v_2 - v_1^2 = 2 v_1 v_2 - u (v_1 + v_2) + (u - v_1)(v_1 - v_2)>0$ so that $\kappa_{22}^{-1}$ is positive.  
     
For $\beta = 0$, we find that 
        \begin{align}
        &\rho_1(k_1) = 
        \begin{dcases}
        \frac{1}{4\pi} \left(1 - \frac{u - v_1}{u - v_2} \right), &k_1<k_1^f, \\
        0, &k_1>k_1^f,
        \end{dcases}\nonumber\\
        &\rho_2(k_2) = 
        \begin{dcases}
        0, &k_2<k_2^f,\\
        \frac{1}{4\pi} \left(1 + \frac{u - v_2}{u - v_1} \right), &k_2>k_2^f,
        \end{dcases}
        \end{align}
where in this case, both Fermi momenta are given by their values in the non-interacting case        
        \beq
		k_1^f = \frac{h_1}{v_1},\quad k_2^f = \frac{h_2}{v_2}.
		\eeq
The compressibility associated with this case is
        \begin{equation}
        \kappa^{-1} = 4 \pi \left(
        \begin{array}{cc}
        v_1 \frac{u - v_2}{v_1 - v_2} & 0\\
        0 & - v_2 \frac{u - v_1}{2u - v_1 - v_2}
        \end{array} \right).
        \end{equation} 
        
Finally, for $\beta>0$ we find that the behavior again changes, with the momentum densities
        \begin{align}
        &\rho(k_1) = 
        \begin{dcases}
        \frac{1}{4\pi} \left(1 - \frac{u - v_1}{u - v_2} \right), &k_1<k_1^{(b)}, \\
        0, &k_1>k_1^{(b)},
        \end{dcases}\nonumber\\
        &\rho_2(k_2) = 
        \begin{dcases}
        0, &k_2<k_2^f,\\
        \frac{1}{2\pi}, &k_2^f<k_2<k_2^{(a)},\\
        \frac{1}{4\pi} \left(1 + \frac{u - v_2}{u - v_1} \right), &k_2>k_2^{(a)},
        \end{dcases}
        \end{align}
such that the Fermi momenta in this case are        
        \beq
		k_1^f = k_1^{(b)},\quad k_2^f =\frac{h_2}{v_2},
		\eeq
leading to
        \begin{equation}
        \kappa^{-1} = 2 \pi \left(
        \begin{array}{cc}
        2 u - v_2 & v_2\\
        v_2 & -v_2
        \end{array} \right).
        \end{equation}
        
For this case, we see that for $\beta<0$, the $v_1$ particles are non-interacting near their Fermi momentum while the $v_2$ particles are interacting throughout their Fermi sea. For $\beta>0$ the roles are reversed since the $v_1$ fermions now interact throughout their Fermi sea while the $v_2$ particles become non-interacting in the vicinity of their Fermi momenta. The discontinuity in $\kappa$ as $\beta$ is changed captures precisely the change in the behavior of the fermions in the vicinity of their respective Fermi momenta.

        \subsubsection{Case A2ii: $u>v_1>0>v_2$ \& $u > \frac{2 v_1 v_2}{v_1 + v_2}$}  
        \label{sec:A2b}
        
The parameter $\beta$ in this case is simply related to the density $D_2$ of the $v_2$ fermions,        
        \begin{equation}
        \beta(D_1,D_2) = D_2.
        \end{equation}
Hence, in this case the regime where $\beta<0$ is un-physical and we only consider the change in $\beta$ as it acquires a non-vanishing value. The excitation spectra are shown in Fig.~\ref{fig:03}.     

The case $\beta = 0$ corresponds to a vanishing density of $v_2$ particles and thus we expect that the $v_1$ particles should exhibit non-interacting behavior throughout the Fermi sea. This is indeed the case, since 
        \begin{align}
        &\rho_1(k_1) = 
        \begin{dcases}
        \frac{1}{2\pi}, &k_1<k_1^f, \\
        0, &k_1>k_1^f,
        \end{dcases} \nonumber\\
        &\rho_2(k_2) = 
        \begin{aligned}
        0 \quad \forall k_2,
        \end{aligned}
        \end{align}
where the Fermi momentum $k_1^f = \frac{h_1}{v_1}$ equals its value in the non-interacting case. Moreover, the compressibility $\kappa_{11}$ also matches that for non-interacting chiral fermions
        \begin{equation}
        \kappa^{-1} = 2 \pi \left(
        \begin{array}{cc}
        v_1 & 0\\
        0 & 0
        \end{array} \right).
        \end{equation}
                    
Once we have a finite density of $v_2$ particles, $\beta = D_2 >0$, and the ground state densities become 
        \begin{align}
        &\rho_1(k_1) = 
        \begin{dcases}
        \frac{1}{2\pi}, &k_1<k_1^{(a)}, \\
        \frac{1}{4\pi} \left(1 - \frac{u - v_1}{u - v_2} \right), &k_1^{(a)}<k_1<k_1^{(b)},\\
        0, &k_1>k_1^{(b)},
        \end{dcases} \nonumber\\
        &\rho_2(k_2) = 
        \begin{dcases}
        0, &k_2<k_2^{f,1},\\
        \frac{1}{2\pi}, &k_2^{f,1}<k_2<k_2^{(a)}, \\
        \frac{1}{4\pi} \left(1 + \frac{u - v_2}{u - v_1} \right), &k_2^{(a)}<k_2<k_2^{(b)},\\
        0, &k_2>k_2^{(b)}.
        \end{dcases}
        \end{align}
Here, the fermions with velocity $v_1$ have a single Fermi momentum while the spectrum for the other species crosses zero energy twice, resulting in two Fermi momenta for that species,
		\beq
		k_1^f = k_1^{(b)},\quad k_2^{f,1} = \frac{h_2}{v_2},\quad k_2^{f,2} = k_2^{(b)}.
		\eeq        
        
The associated compressibilities in this case are
        \begin{equation}
        \kappa^{-1} = 2 \pi \left( 
        \begin{array}{cc}
         v_1 & -v_2 \frac{u - v_1}{u - v_2} \\
         -v_2 \frac{u - v_1}{u - v_2} & - v_2 \frac{u^2 - v_1 v_2}{(u - v_2)^2}
        \end{array} \right).
        \end{equation}
Here, the $v_1$ particles interact near the Fermi sea while the $v_2$ particles develop two Fermi momenta as a consequence of their spectrum $\epsilon_2(k_2)$ becoming non-chiral. 
        
        \subsubsection{Case A3: $u>0>v_1>v_2$}
        \label{sec:A3}
        
\begin{figure}[t]
    \includegraphics[width=8cm]{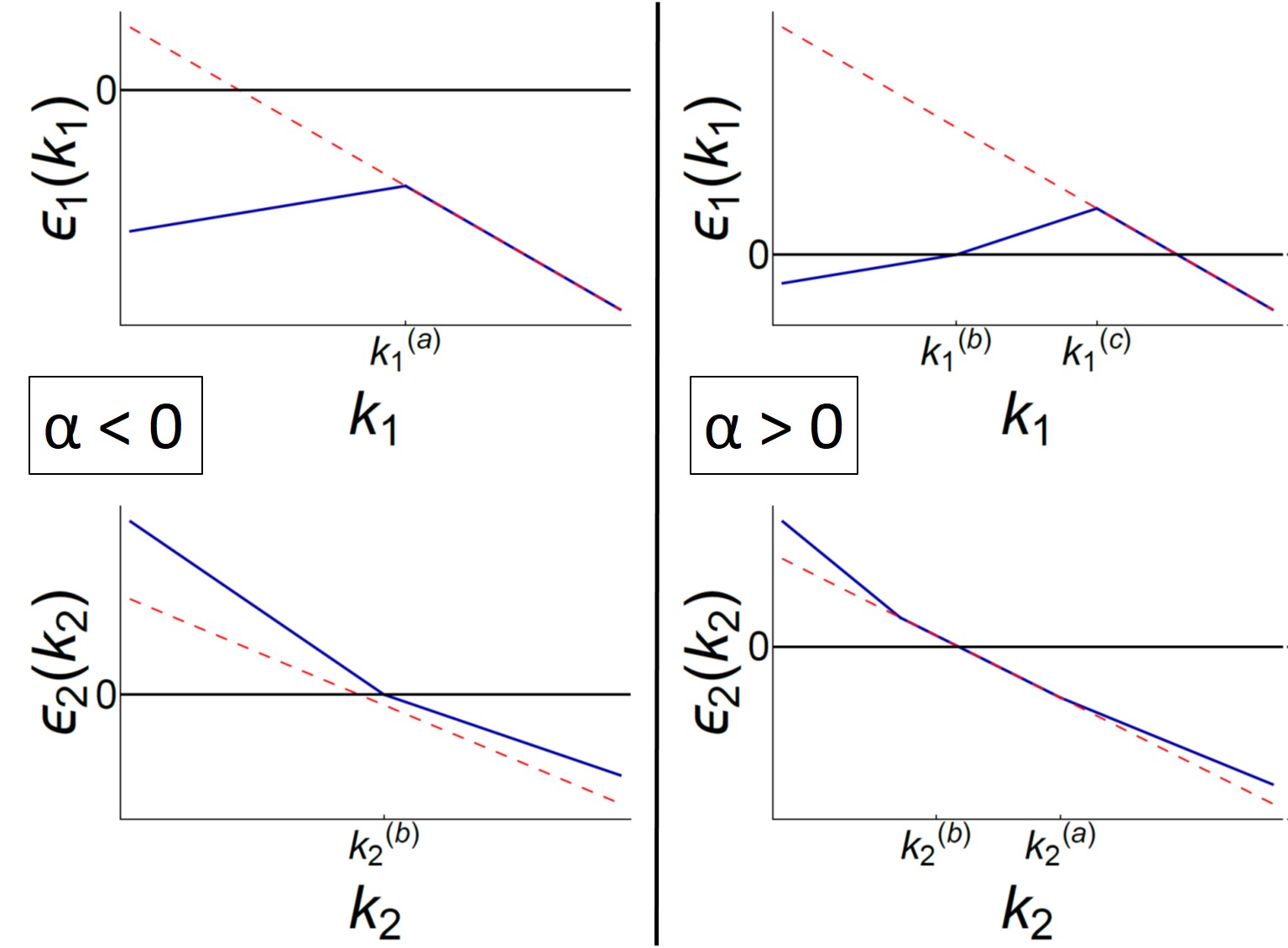}
    \centering
    \caption{\label{fig:07} Excitation spectra for case A3, with $u = 8,v_1 = -3,v_2 = -6,h_1 = 1,h_2 = 2$. The plots on the left are for $\alpha<0$ ($\nu = -10$) in which case the $v_1$ fermions do not develop a Fermi momentum. The right-hand panel corresponds to $\alpha>0$ ($\nu = 30$) where the $v_1$ particles develop two distinct Fermi momenta and where the $v_2$ particle are interacting deep into the Fermi sea. The red dashed lines are the excitation spectra for a non-interacting system ($g = 0$).}
    \end{figure}

For this case, we find that 
        \begin{equation}
        \beta(D_1,D_2) = \frac{\nu  \left(2 u-v_1-v_2\right)}{\left(u-v_1\right) \left(u-v_2\right)}-4 \pi  \left(D_1+D_2\right).
        \end{equation}        
Here, we find that $\beta<0$ implies the lack of any Fermi momentum for the $v_1$ particles i.e., $\epsilon_{\rho}(k)<0\, \forall\,k$ as depicted in the left panel of Fig.~\ref{fig:07}. Since there is no Fermi momentum, we cannot find a relation between the particle densities $D_i$ and the chemical potentials $h_i$, such that the compressibility is an ill-defined quantity in this case.   

However, for $\beta = 0$ the $v_1$ particles develop a Fermi momentum and we find that 
        \begin{align}
        &\rho_1(k_1) = 
        \begin{dcases}
        \frac{1}{4\pi} \left(1 - \frac{u - v_1}{u - v_2} \right), &k_1<k_1^f ,\\
        \frac{1}{2\pi}, &k_1>k_1^f,
        \end{dcases} \nonumber\\
        &\rho_2(k_2) = 
        \begin{dcases}
        0, &k_2<k_2^f,\\
        \frac{1}{4\pi} \left(1 + \frac{u - v_2}{u - v_1} \right), &k_2>k_2^f,
        \end{dcases}
        \end{align}
where the Fermi momenta take on their non-interacting values,
\beq
k_1^f = \frac{h_1}{v_1},\quad k_2^f = \frac{h_2}{v_2},
\eeq            
from which we find
        \begin{equation}
        \kappa^{-1} = -\frac{4 \pi}{2u - v_1 - v_2} \left(
        \begin{array}{cc}
        v_1 (u - v_2) & 0\\
        0 & v_2 (u - v_1)
        \end{array} \right).
        \end{equation}   
    
For $\beta>0$, we find that the behavior changes and, similar to the previous case Sec.~\ref{sec:A2b}, we find that the $v_1$ fermions no longer have a fixed chirality due to the weak interactions. Furthermore, the $v_2$ particles remain interacting deep within the Fermi sea and it is only in the vicinity of their Fermi momentum that they do not interact. This is reflected in the ground state densities, 
        \begin{align}
        &\rho_1(k_1) = 
        \begin{dcases}
        \frac{1}{4\pi} \left(1 - \frac{u - v_1}{u - v_2} \right), &k_1<k_1^{(b)}, \\
        0, &k_1^{(b)}<k_1<k_1^{f,1} ,\\
        \frac{1}{2\pi}, &k_1>k_1^{f,1},
        \end{dcases} \nonumber\\
        &\rho_2(k_2) = 
        \begin{dcases}
        0, &k<k_2^f,\\
        \frac{1}{2\pi}, &k_2^f<k_2<k_2^{(a)} ,\\
        \frac{1}{4\pi} \left(1 + \frac{u - v_2}{u - v_1} \right), &k_2>k_2^{(a)}.
        \end{dcases}
        \end{align}
Here the first species of fermions, with velocity $v_1$, develop two Fermi momenta since their spectrum crosses zero energy twice, while the second species of fermions have a Fermi momentum that takes on its non-interacting value,
\beq
k_1^{f,1} = \frac{h_1}{v_1},\quad k_1^{f,2} = k_2^{(b)},\quad k_2^f = \frac{h_2}{v_2}.
\eeq    
From these ground state densities, we derive 
        \begin{equation}
        \kappa^{-1} = -\frac{2 \pi}{2u - v_1 - v_2} \left(
        \begin{array}{cc}
        v_1 (2 u - v_2) & v_1 v_2\\
        v_1 v_2 & v_2 (2u - v_1)
        \end{array} \right).
        \end{equation}
        
As in all previous cases, the discontinuity in the compressibilities $\kappa$ indicates the phase transition that occurs in the Q3WI model as a function of the particle densities $D_i$ and the detuning $\nu$. In addition to the $v_1$ particles being non-chiral for $\beta>0$, this case exhibits the peculiar behavior that when $\beta < 0$, the $v_1$ fermions are unable to develop a Fermi momentum, with $\kappa$ ill-defined in this regime.          
        
        \subsubsection{Case A4: $0>u>v_1>v_2$}
        \label{sec:A4}
        
This is the final region of parameter space which appears in case A. Here, we find that 
        \begin{equation}
        \beta(D_1,D_2) = \nu -2 \pi  D_1 \left(u-v_1\right)-2 \pi  D_2 \left(u-v_2\right).
        \end{equation}
           
  	\begin{figure}[t]
    \includegraphics[width=8cm]{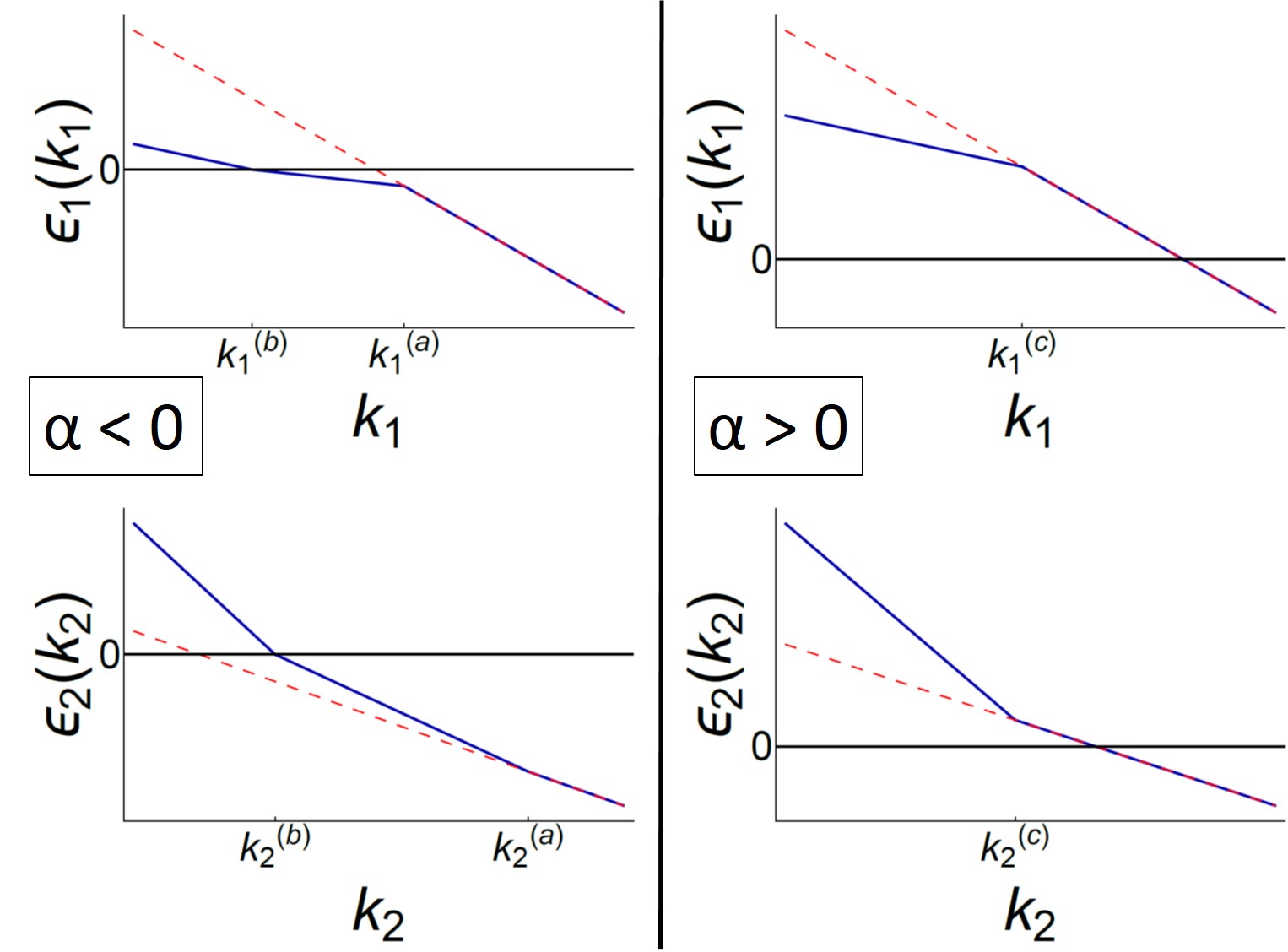}
    \centering
    \caption{\label{fig:08} Excitation spectra for case A4, with $u = -2,v_1 = -4,v_2 = -10,h_1 = 1,h_2 = 2$. The plots on the left are for $\nu = -10$, so for $\alpha<0$, while the ones on the right are for $\nu = 10$, so for $\alpha>0$. The red dashed lines are the excitation spectra for a non-interacting system.}
    \end{figure} 
       
    For $\beta<0$, we find that both species of fermions remain non-interacting deep within the Fermi sea but interact in the vicinity of their Fermi momenta, as shown in Fig.~\ref{fig:08} and reflected in the ground state densities    
        \begin{align}
        &\rho_1(k_1) = 
        \begin{dcases}
        0, &k_1<k_1^{(b)}, \\
        \frac{1}{4\pi} \left(1 - \frac{u - v_1}{u - v_2} \right), &k_1^{(b)}<k_1<k_1^{(a)}, \\
        \frac{1}{2\pi}, &k_1>k_1^{(a)},
        \end{dcases}\nonumber\\
        &\rho_2(k_2) = 
        \begin{dcases}
        0, &k_2<k_2^{(b)},\\
        \frac{1}{4\pi} \left(1 + \frac{u - v_2}{u - v_1} \right), &k_2^{(b)}<k_2<k_2^{(a)},\\
        \frac{1}{2\pi}, &k_2>k_2^{(a)},
        \end{dcases}
        \end{align}
with the Fermi momenta
\beq
k_1^f = k_1^{(b)},\quad k_2^f = k_2^{(b)}, 
\eeq
and associated compressibilities,
        \begin{equation}
        \kappa^{-1} = \frac{-2 \pi}{2u - v_1 - v_2} \left(
        \begin{array}{cc}
        u^2 - v_1 v_2 & (u - v_1)(u - v_2)\\
        (u - v_1)(u - v_2) & u^2 - v_1 v_2
        \end{array} \right).
        \end{equation}
        
The behavior simplifies for $\beta\geq 0$ where we find that both species of fermions remain completely non-interacting throughout the Fermi sea,
        \begin{align}
        &\rho_1(k_1) = 
        \begin{dcases}
        0, &k_1<k_1^f ,\\
        \frac{1}{2\pi}, &k_1>k_1^f,
        \end{dcases}\nonumber\\
        &\rho_2(k_2) = 
        \begin{dcases}
        0, &k_2<k_2^f,\\
        \frac{1}{2\pi}, &k_2>k_2^f,
        \end{dcases}
        \end{align}
with Fermi momentum $k_i^f = h_i/v_i$ and compressibility
        \begin{equation}
        \kappa^{-1} = -2 \pi \left(
        \begin{array}{cc}
        v_1  & 0\\
        0 & v_2
        \end{array} \right),
        \end{equation}
also corresponding to that of non-interacting chiral fermions. We note that qualitatively this case is similar in nature to case A1, which is discussed in the main text.     
        
Thus, in each of the preceding cases, we have demonstrated that the compressibilites $\kappa$ display a discontinuity as a function of $D_1,D_2,\nu$ since the ground state densities $\rho_i(k_i)$ are discontinuous. We further note that in each case $\kappa_{ii}$ (i = 1,2) are positive, as required on physical grounds. We see that det$(\kappa)>0$ in each case as well.

	\section{Solution to Integral Equations}
	\label{app:E}
Here, we discuss the procedure for finding a solution to an integral equation of the form encountered in the main text, Eq.~\eqref{eqn:129}. To do this, we us consider a more general integral equation of the form
		\begin{equation}
		\int_{a}^b d\mu\,\frac{\phi(\mu)}{(\mu - \lambda)^2} = f(\lambda) ,\quad \lambda \in (a,b).
		\end{equation}				
Then, we define 
		\begin{equation}
		\mu = \frac{b - a}{2}x + \frac{b + a}{2} \, ,\quad \lambda = \frac{b - a}{2}y + \frac{b + a}{2} ,
		\end{equation}	
such that $x,y\in[-1,1]$ and $\mu - \lambda = \frac{b - a}{2} (x-y)$. This leads to 
		\begin{equation}
		\int_{-1}^1 dx\,\frac{\phi(\frac{b - a}{2}x + \frac{b + a}{2})}{(x - y)^2} = \frac{b - a}{2} f\left(\frac{b - a}{2}y + \frac{b + a}{2}\right) ,
		\end{equation}
or, equivalently
		\begin{equation}
		\int_{-1}^1 dx\,\frac{A(x)}{(x-y)^2} = B(y) , \quad y \in [-1,1] ,
		\end{equation}
where $A(x) = \phi(\mu)$ and $B(y) = \frac{b - a}{2} f(\lambda)$. 

From the general theory of hyper-singular equations, (see, e.g. Refs.~\cite{Muskh, Lifanov}), the solution to the above equation is known to be
		\begin{align}
		A(y) &= \int_{-1}^{y} \frac{d\tau}{\sqrt{1 - \tau^2}} \int_{-1}^1 \, dx \frac{\sqrt{1 - x^2}}{\tau - x} \frac{B(x)}{\pi^2} ,\nonumber \\
		&= \int_{-1}^1 \, dx\,\frac{B(x)}{\pi^2} \log\left[ \frac{\sqrt{(1 - x)(1 + y)} - \sqrt{(1 + x)(1 - y)}}{\sqrt{(1 - x)(1 + y)} + \sqrt{(1 + x)(1 - y)}} \right].
		\end{align}
For instance, for $B(y) = C$, a constant, we find
		\begin{equation}
		A(x) = -\frac{C}{\pi}\sqrt{1 - x^2} ,\quad x\in [-1,1],
		\end{equation}				
while for $B(y) = C y$, we find
		\begin{equation}
		A(x) = -\frac{C}{2\pi} x \sqrt{1 - x^2} ,\quad x\in [-1,1].
		\end{equation}
		

	\section{Compressibilities in the Quasi-Condensate regime}
	\label{app:F}		

In this Appendix, we discuss in detail the calculation establishing the ground state compressibility~\eqref{eq:ccco} for the quasi-condensate regime (see Sec.~\ref{condensate}). 

Specifically, let us consider the case where $v_1>u>v_2>0$ and where the parameter $\alpha<0$. The behavior of the fermions in this regime is depicted in Fig.~\ref{fig:04} and the ground state densities to leading order are given by Eqs.~\eqref{rhononint},~\eqref{rho1}, and~\eqref{rho2}. For ease of notation, we denote the Fermi momenta throughout this section as
\beq
K_i \equiv k_i^f.
\eeq
In what follows, we will require knowledge of the matrix of derivatives
\beq
M_{ij} = \frac{\p K_i}{\p D_j} = (V^{-1})_{ij},
\eeq
where the matrix $V$ is defined as
\beq
V_{ij} = \frac{\p D_i}{\p K_j}.
\eeq

From Fig.~\ref{fig:04} and the expressions for $\rho_i$, we see that the density of the first fermionic species $D_1$ depends non-trivially on the Fermi momentum of both species of fermions since the parameter $k_1^{(i)}$ is a function of $K_2$ (see Eq.~\eqref{kint}). However, the density of the other species $D_2$ depends only on its own Fermi momentum $K_2$, so that 
	\beq
	\frac{\p D_2}{\p K_1} = 0.
	\eeq	
We also note that while the particle densities and total energy diverge in the limit $g \to 0$, their derivatives may remain finite. For instance, the only dependence of $D_1$ on $K_1$ comes from the range of momenta over which $\rho_1$ takes its non-interacting value, from which we can show that 
	\beq
	\frac{\p D_1}{\p K_1} = \frac{1}{2 \pi} + O(g^2).
	\eeq  
Similarly,	
	\beq
	\frac{\p^2 \mathcal{E}}{\p K_1^2} = \frac{v_1}{2 \pi} + O(g^2).
	\eeq
We hence find that the expressions for the compressibilities simplify considerably. For the inverse compressibility of the first species of fermions we find that 
	\beq
	[\kappa]_{11}^{-1} = \left( \frac{\partial D_1}{\partial K_1}\right)^{-2} \frac{\partial^2 \mathcal{E}}{\partial K_1^2} = 2 \pi v_1 + O(g^2),
	\eeq
which reflects the fact that $\rho_1$ takes its non-interacting value of $1/(2\pi)$ in the vicinity of $K_1$.
	
Next, we consider the mixed term	
	\beq
	[\kappa]_{12}^{-1} = -2 \pi v_1 \left(\frac{\p D_1}{\p K_2} \right) \left(\frac{\p D_2}{\p K_2} \right)^{-1} = [\kappa]_{21}^{-1}.
	\eeq
Crucially, we see from Fig.~\ref{fig:04} that the dependence of $D_1$ ($D_2$) on $K_2$ comes through the region where $\rho_1$ ($\rho_2$) takes its interacting value and since we know that our analytic solution for $\rho_1$ breaks down in the vicinity of $K_2$, care must be taken when evaluating a term such as $\p D_1/\p K_2$. In particular, we must account for the possibility that corrections to our analytic solution alter the leading order solution for $\kappa^{-1}$. This issue does not arise for our solutions in the non-interacting regions since any corrections to the $1/(2\pi)$ density in those regions vanish as $g\to 0$. Since our analytic solutions within the interacting regions $\sim 1/g^2$ however, corrections to those may still diverge in the limit $g\to 0$ and hence must be treated more carefully.
	
In order to take into account the effect of corrections to our analytic solution within the interacting regions, let us consider 
\beq
\frac{\p D_1}{\p K_2} = \frac{\p}{\p K_2} \left[\int_{-\Lambda}^{k_1^{(i)}}dk_1\,\tilde{\rho}_1(k_1) + \frac{1}{2 \pi} \int_{k_1^{(i)}}^{K_1} dk_1\right],
\eeq	
where $k_1^{(i)}$ depends on $K_2$ and where 
\beq
\tilde{\rho}_1 = \rho_1 + \delta \rho_1,
\eeq
with $\delta \rho_1$ representing corrections to our leading order solution $\rho_1$ in the interacting region, given by Eq.~\eqref{rho1}. We can further simplify the above expression to find 
\beq
\frac{\p D_1}{\p K_2} = \left(\tilde{\rho}_1(k_1^{(i)}) - \frac{1}{2\pi} \right) \frac{d k_1^{(i)}}{\p K_2} + \int_{-\Lambda}^{k_1^{(i)}} dk_1\, \frac{\p}{\p K_2}\tilde{\rho_1}(k_1).
\eeq
Now, while our analytical solution Eq.~\eqref{rho1} for $\rho_1$ vanishes at $k_1^{(i)}$, we know that this is merely an artefact of the small $g$ approximation and that the real solution should connect smoothly with $1/(2\pi)$. Thus, we expect that
\beq
\frac{\p D_1}{\p K_2} = \int_{-\Lambda}^{k_1^{(i)}} dk_1\, \frac{\p}{\p K_2}\tilde{\rho_1}(k_1),
\eeq
where the dependence of $\tilde{\rho}_1$ on $K_2$ is implicit. 

A similar analysis demonstrates that
\beq
\frac{\p D_2}{\p K_2} = \tilde{\rho}_2(K_2) + \int_{-\Lambda_0}^{K_2} dk_2\, \frac{\p}{\p K_2}\tilde{\rho_2}(k_2),
\eeq	
with 
\beq
\tilde{\rho}_2 = \rho_2 + \delta \rho_2,
\eeq
where $\delta \rho_2$ represents corrections to our leading order solution $\rho_2$ in the interacting region, given by Eq.~\eqref{rho2}. Again, we note that our leading order solution vanishes exactly at $K_2$ i.e., $\rho_2(K_2) = 0$, but we expect that the true solution will remain finite at the end point, in analogy with the Lieb-Liniger gas. 
	
Importantly, we note that since our leading order analytic solutions $\sim 1/g^2$, any corrections to these must be less singular i.e., $\delta \rho_i \sim 1/g^\alpha$ with $\alpha<2$. Consequently, any corrections to the derivatives $\p D_i/\p K_2$ must also be less singular and we thus find that the ratio 
\beq
\left( \frac{\p D_1}{\p K_2}\right) \left(\frac{\p D_2}{\p K_2} \right)^{-1} = 1 + O(g)
\eeq
in the limit $g \to 0$, since our leading order solutions for the ground state densities satisfy
\beq
\int_{-\Lambda}^{k_1^{(i)}} dk_1\, \frac{\p}{\p K_2}\rho_1(k_1) = \int_{-\Lambda_0}^{K_2} dk_2\, \frac{\p}{\p K_2}\rho_2(k_2),
\eeq
and because all contributions from the corrections $\delta \rho_i$ vanish as $g\to 0$. We have hence established that it is sufficient to consider our analytic expressions for the ground state densities, derived in the main text, in order to find the leading order compressibilities. Following the above discussion, we find that 
\beq
[\kappa]_{12}^{-1} = -2 \pi v_1 + O(g) = [\kappa]_{21}^{-1}.
\eeq
Similarly, we can establish that
\beq
[\kappa]_{22}^{-1} = 2 \pi v_1 + O(g).
\eeq

To demonstrate that the inverse compressibility matrix $\kappa^{-1}$ is invertible at next-to-leading order, we note that 
\beq
[\kappa]_{22}^{-1} = \frac{\p^2 \mathcal{E}}{\p D_2^2} = \frac{1}{2\pi v_1} [\kappa]_{12}^{-2} + \Gamma,
\eeq
where
\begin{align}
\Gamma &= \left(\frac{\p D_2}{\p K_2}\right)^{-2} \frac{\p^2 \mathcal{E}}{\p K_2^2} \nonumber \\
&+ \left(\frac{\p D_2}{\p K_2} \right)^{-1} \frac{\p \mathcal{E}}{\p K_2} \frac{\p}{\p K_2} \left[\left(\frac{\p D_2}{\p K_2}\right)^{-1}\left(1 - 2 \pi \frac{\p D_1}{\p K_2} \right) \right].
\end{align}

Following the above discussion, we have seen that we can write 
\beq
[\kappa]_{12}^{-1} = -2 \pi v_1 (1 + \epsilon_1)
\eeq	
where $\epsilon_1 \to 0$ as $g \to 0$. Similarly, we can evaluate the behavior of $\Gamma$ while taking into account corrections to the leading order behavior of $\rho_i$ and we find that 
\beq
\Gamma = 2 \pi v_1 \epsilon_2,
\eeq
where $\epsilon_2 \to 0$ as $g \to 0$ i.e., there is no finite contribution to $\Gamma$ from the leading order solution in the limit of small $g$. We can now write 
\beq
[\kappa]_{22}^{-1} = 2 \pi v_1 [(1 + \epsilon_1)^2 + \epsilon_2] \approx 2\pi v_1(1 + 2\epsilon_1 + \epsilon_2)
\eeq
as $g \to 0$. Thus, the inverse compressibility matrix, accounting for the leading order corrections is 
\beq
\kappa^{-1} = \left( 
\begin{array}{cc}
2\pi v_1 & -2\pi v_1 (1 + \epsilon_1)\\
-2\pi v_1 (1 + \epsilon_1) & 2\pi v_1 (1 + 2\epsilon_1 + \epsilon_2) \end{array}
\right),
\eeq
with a determinant given by
\beq
\det(\kappa^{-1}) = (2 \pi v_1)^2 \epsilon_2
\eeq
to leading order in $g$. Hence, we see that that non-invertible nature of $\kappa^{-1}$ is a consequence of the small $g$ approximation and that it is necessary to establish corrections to the analytical solutions developed in this paper in order to find the compressibility matrix $\kappa$. While we have discussed the specific case when $\alpha<0$ here, the case when $\alpha>0$ can be analyzed similarly.


\newpage 

\bibliography{library}


\end{document}